\documentclass[prd,aps,nofootinbib,preprint]{revtex4}
\usepackage{amsmath,amssymb,amsbsy,bm,amsfonts,color,graphicx,graphics,latexsym,placeins,epsfig,multirow}
\usepackage{array}
\usepackage{tikz}
\newcolumntype{P}[1]{>{\centering\arraybackslash}p{#1}}
\usepackage{comment}
\usepackage{footmisc}
\usepackage[compatibility=false]{caption}
\usepackage{subcaption}
\usepackage{bbold}
\usepackage{enumitem}
\setdescription{leftmargin=12.5pt}
\usepackage{bbold}
\usepackage{hyperref}

\usepackage{natbib}
\setcitestyle{square, comma, numbers, sort&compress}
\usepackage{float}
\newcommand*{\id}{{\rm\hbox{1\kern-0.15em \vrule width .1pt depth-.2pt}}}

\begin{document}


\title{\Large \bf A simple analytic example of the gravitational wave memory effect}
\author{Indranil Chakraborty and
Sayan Kar}
\email{indradeb@iitkgp.ac.in,   sayan@phy.iitkgp.ac.in}
\affiliation{Department of Physics \\ Indian Institute of Technology Kharagpur, 721 302, India}

\begin{abstract}
 
\noindent We report an analytical example of the gravitational wave memory effect in exact plane wave spacetimes. A square pulse profile is chosen which gives rise to a curved wave region sandwiched between two flat Minkowski spacetimes. Working in the Brinkmann coordinate system, we solve the geodesic equations exactly in all three regions. Issues related to the continuity and differentiability of the solutions at the boundaries of the pulse are addressed. The evolution of the geodesic separation reveals displacement and velocity memory effects 
with quantitative estimates depending on initial values and the amplitude, width of the pulse. The deformation caused by the pulse on a ring of particles is then examined 
in detail. Formation of caustics is found in both scenarios i.e. 
evolution of separation for a pair of geodesics and
shape deformation of a ring of particles -- a feature 
consistent with previous work on geodesic congruences in this 
spacetime. In summary, our analysis provides a useful illustration of memory effects involving closed-form exact expressions. 
 
\end{abstract}

\maketitle

\section{Introduction}

\noindent A unique nonlinear feature of gravitational waves is that it leaves an imprint over the spacetime it passes through \citep{Favata:2010}. This phenomena is known as the gravitational wave memory effect and it causes a permanent change in the separation between freely falling test particles \citep{Favata:2010,Bieri:2017arXiv}. It is one of the strong field effects of General Relativity (GR) that remains undetected, till date, in astrophysical  observations \citep{Hubner:2021,Aggarwal:2019}. 

\noindent Gravitational memory effects were earlier  classified into linear \citep{Zeldovich:1974,Braginsky:1985} and nonlinear effects \citep{Christodoulou:1991} 
based on the type of theory used to study it. In the linear memory effect it was shown that in case of hyperbolic scattering, the metric perturbation before and after the event changes \citep{Zeldovich:1974}. This was found to be related to the difference in the quadrupole moment of the radiation at late and early times \cite{Braginsky:1985}. On the nonlinear side, Christodoulou, using full GR, showed how gravitational waves travelling to null infinity gives rise to an additional contribution to memory. 
Recent work \citep{Bieri_gaugeinv:2014} using gauge invariant observables have shown that this classification is a misnomer and one can find both types of memory effects in linearized gravity. Subsequently, they have been identified as the ordinary  and null memory effects corresponding to  gravitational radiation sourced by massive and massless particles respectively \citep{Tolish1:2014}. Similar works in this direction appear in \citep{Tolish:2014, Madler:2016,Madler:2017}. Moreover, an interesting theoretical connection has been conjectured between the memory effects, the Bondi-van der Burg-Metzner-Sachs (BMS) symmetries and soft theorems \citep{Strominger:2016,Strominger:2017}. It is shown how the gravitational 
 memory effect can be realized in asymptotically flat spacetimes as a transition between two degenerate Minkowski vacua having different numbers of soft hair \citep{Compere:2019}. Such relationships in infrared physics have been investigated, apart from GR, for Brans-Dicke gravity \citep{HouJHEP:2020,Hou1:2020,Tahura:2021} and Chern-Simons theory \citep{Hou1:2021}. 

\noindent In this work, we consider a much simplified scenario consisting of an exact plane wave spacetime \citep{Bondi:1959,Peres:1959}. Exact plane gravitational waves are nonlinear vacuum solutions of Einstein Field Equations. There already exists considerable amount of literature on memory effects in this spacetime \citep{Zhang:2017,Zhang:2017soft,Zhang:2018vel,Zhang:2018,Zhang:2018SL,Chak:2020,Cvetkovic:2021}. Our current work is motivated from an analysis done in \citep{Zhang:2017soft} where the authors demonstrate memory effects by studying geodesic evolution. In this geometry, there are {\em free functions} in the metric line element which 
may be used to appropriately
define a pulse profile in the gravitational wave. The authors in
\citep{Zhang:2017soft} carried out the analysis by solving the geodesic equations {\em numerically} for a Gaussian pulse profile. Parallel comoving geodesics were shown to exhibit monotonically increasing separation after encountering the pulse, thereby showing both displacement (change in the separation) and velocity (change in relative velocity) memory effects. The present work provides a complete {\em analytic} solution of the memory effect in the exact plane wave spacetime, {\em albeit} with a different pulse profile. To achieve this, we consider a {\em square pulse} for our calculation, which, not only leads to exact solvability but is
also representative of the basic physics related to memory. 

\noindent  Using the same pulse profile, earlier work on geodesic congruences by the current authors have shown that gravitational waves cause timelike geodesic congruences to focus while the shear is found to diverge \citep{Chak:2020}. The focusing depends on the amplitude and width of the pulse. This type of memory effect is called ${\cal B}$-memory in the literature \citep{Loughlin:2019,Srijit:2019}. As mentioned earlier, in this paper, we work out displacement and velocity memory effects for geodesics traversing this spacetime. Though in ${\cal B}$-memory we study the gradient of the velocity field, it is not the same as velocity memory.

\noindent A square pulse profile in an exact plane gravitational wave spacetime gives rise to a {\em sandwich wave geometry}. Such a geometry has two flat Minkowski spacetimes separated by a region containing gravitational waves. We solve geodesic equations analytically in all the three regions separately. The geodesic solutions along transverse directions are assumed to be continuous and differentiable along the boundaries of the pulse. The evolution of the geodesic separation is noted in all the three regions. We, at first, examine separate cases of plus and cross polarization and then study the case having both polarizations to be non-zero. In all the cases, we find monotonic displacement memory and constant shift velocity memory effects.  Furthermore, a separate analysis is carried out for understanding the effects of gravitational wave memory on a ring of particles. The deformation in the configuration is shown to be related to the shape of the pulse. Relevant kinematical variables are computed and the results are shown to be consistent with results obtained from earlier work \citep{Harte:2012,Chak:2020}. 
Finally, a brief note on the memory effect obtained using geodesic deviation is also provided here. The two methods (geodesic and deviation) are shown to yield identical results.

\noindent The organization of the paper is as follows. In Section II, we discuss exact plane wave spacetimes. Section II A describes the various coordinate systems used for studying gravitational plane wave. In Section II B, we focus on the sandwich wave geometry. Section II C gives the geodesic equations in this spacetime. Memory effects inferred from studying the geodesic equations for only plus and cross polarization  are given in Section III A and B respectively. The analysis for a ring of particles is also provided under the same respective sections. Memory effects when both polarizations are simultaneously present is discussed in Section III C. Section III D  touches upon memory effects using the geodesic deviation equation. A brief summary of the results presented in this paper is given in Section IV. 
\vspace{-0.2in}
\section{Exact plane wave spacetimes}
 
\noindent The family of {\em pp}-wave (plane fronted waves with parallel rays) spacetimes are defined to have a covariantly constant null vector field \citep{Stephani:2003,Griffiths:2009}. Exact plane wave spacetimes fall within the class of general {\em pp}-wave spacetimes where the Riemann curvature is constant over each wavefront \citep{Griffiths:2009}. There are two standard coordinate systems in which the line element may be written\textemdash (a) Baldwin-Rosen-Jeffrey (BJR) coordinates \citep{Rosen:1937}  and (b) Brinkmann coordinates \cite{Brinkmann:1925}. 
We first discuss each of the coordinate systems  and state the reason for choosing the Brinkmann coordinates in the ensuing calculations on memory effects. Thereafter, we describe the sandwich wave spacetime geometry used in this work followed by a discussion on the geodesic equations in such spacetimes.

\subsection{Coordinate systems}
\vspace{-0.1in}
\subsubsection{BJR coordinates}

\noindent The line element in BJR coordinates is given below.
\begin{align}\label{eq:BJR}
    ds^2=-2dudV+a_{ij}(u)dX^idX^j
\end{align}
\noindent Here, $u$ is the retarded null coordinate. Considering $u=t-z$ and $V=t+z$, we find that the curvature disturbances propagate along the $z$-direction with the speed of light.  The spacetime gets distorted in the 
space orthogonal ({\em i.e.} along $X^1, X^2$) to the direction of propagation. 
Constant $u$ hypersurfaces correspond to planar wavefronts. Hence, the above line element represents a plane gravitational wave spacetime. The function $a_{ij}(u)$ encodes the gravitational wave field. If $a_{ij}=\mathbb{1}_{2\times2}$, the full metric is simply Minkowskian. Moreover, in the transverse traceless (TT) gauge, linearized plane waves can be written down in a perturbative form in this coordinate system, i.e. 
\begin{equation}
    a_{ij}=\eta_{ij}+h_{ij}
\end{equation}
Thus, one may say, that the exact plane wave spacetimes (as given in Eq.(\ref{eq:BJR}))
in full, nonlinear GR, are, in some sense,  generalizations of the linearized gravitational plane waves expressed in the transverse traceless (TT) gauge.

\noindent  In BJR coordinates, Ricci flatness of plane waves spacetimes results in a Sturm-Liouville equation like \citep{Zhang:2017soft},
\begin{equation}
    \ddot{\chi}+\omega^2(u)\chi=0 \label{eq:det_BJR}
\end{equation}

\noindent Here, $\chi=det(a_{ij})^{1/4}$ and $\omega^2(u)=\dfrac{1}{8}\mathrm{Tr}[(\bm{\gamma}^{-1} \dot{\bm{\gamma}})^2]$ where $\gamma_{ij}=\chi^{-2} a_{ij}$. Since $\omega^2(u)>0$, Eq.(\ref{eq:det_BJR}) denotes an attractive oscilattor with time($u$)-dependent frequency. Hence, $\ddot{\chi}<0$ which implies $\chi=0$ at some finite value of $u$. Since the determinant of $a_{ij}$ vanishes at some $u$-value for the plane wave metric, this results in a coordinate singularity. Thus, we do not use BJR coordinates any further, for our calculations in this article.

\vspace{-0.2in}

\subsubsection{Brinkmann coordinates}

 \noindent The line element in the Brinkmann coordinate system is,
 \begin{equation}
ds^2= -H(u,x,y) du^2 -2 dudv + dx^2+dy^2 \label{eq:brinkmann}
\end{equation}
\noindent Coordinate transformations relating BJR to Brinkmann coordinates can be found in \citep{Zhang:2017soft}. Note that the $u$-coordinate is the same in both the coordinate systems.  Consistent with the definition of a {\em pp}-wave, we find that the null vector field $\partial_v$ is covariantly constant ($\nabla.\partial_v=0$).  Here, $u$-constant hypersurfaces are flat due to absence of matter. These wave surfaces are spanned by the vectors $\partial_x$ and $\partial_y$.  For vacuum solutions, the metric function $H(u,x,y)$ satisfies,
\begin{equation}
    H,_{xx}+H,_{yy}=0 \label{eq:laplacian}
\end{equation}
\noindent The general solution \footnote{For Maxwell fields we have a third term in  $H(u,x,y)$ like $B(u)(x^2+y^2)$. } of Eq.(\ref{eq:laplacian}) is
\begin{equation}
H(u,x,y)=\dfrac{1}{2}A_+(u)[x^2-y^2]+A_{\times}(u)xy
\end{equation}
\noindent Such a solution of $H(u,x,y)$ is also consistent with the vacuum wave equation.  The functions $A_+(u)$ (plus),   $A_{\times}(u)$ (cross) denote the pulse profiles of the two polarizations of the gravitational wave field. The nonzero Riemann tensor components then become
\begin{equation}
    R_{xuxu}=\frac{1}{2}A_+(u) \hspace{1.5cm} R_{yuyu}=-\frac{1}{2}A_+(u) \hspace{1.5cm}
    R_{xuyu}=\frac{1}{2}A_\times(u) \label{eq:curvature}
\end{equation}
\noindent From Eq.(\ref{eq:curvature}) we find that the Riemann curvature is same along $u$-constant hypersurfaces (hence the name {\em plane}). They denote planar wavefronts. Since the curvature of the spacetime depends only on the retarded time coordinate $u$ and not on its derivative, one can construct sandwich and shock waves for appropriately chosen pulse profiles  $A_+(u)$ and $A_\times(u)$ \citep{Griffiths:2009}. In our work, we consider such a sandwich wave geometry to understand analytically, the non-linear feature of gravitational waves known as the {\em memory effect}.

\noindent Note that for exact plane waves, all the curvature scalars vanish. Thus, any singularity present in the spacetime is realized by computing the tidal tensor which remains regular for {\em nonsingular} pulse profiles \citep{Blau:2011}. Henceforth, we perform all the calculations using the Brinkmann coordinate system.

\subsection{Sandwich wave spacetime geometry}

\noindent The sandwich wave geometry is easily visualized in BJR coordinates. Consider a curved wave region sandwiched between two flat Minkowski spacetimes. The wave travels along the $V$-direction. The null hyperplanes $u=u_1$ and $u=u_2$ are the boundaries of the curved wave region in this geometry. 
\vspace{-0.2cm}
\begin{figure}[H]
    \centering
    \includegraphics[scale=0.4]{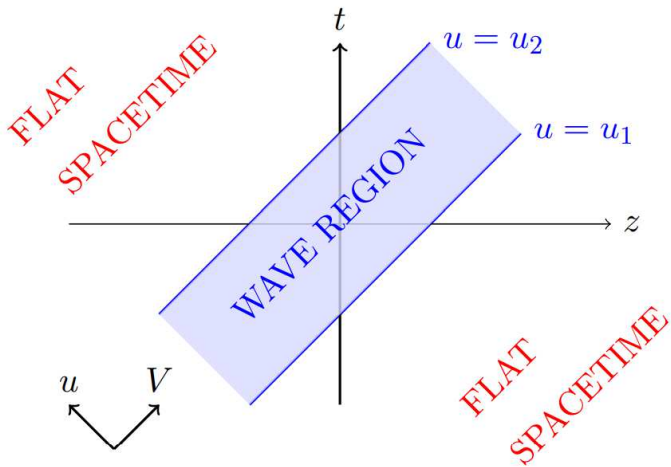}
    \caption{Sandwich wave spacetime construction in BJR coordinates. The wave region is shown in blue. $u_1$ and $u_2$ correspond to wavefronts of the exact plane wave. The flat spacetimes correspond to $\ddot{a}_{ij}(u)=0$ while in the wave region, $\ddot{a}_{ij}(u)\neq 0$.}
    \label{fig:BJR_sandwich_wave}
\end{figure}

\noindent A similar construction in Brinkmann coordinates follows from choosing suitable pulse profiles $A_+(u)$ or $A_\times(u)$. We work with the following analytical form of the pulse profile
\begin{equation}
    A_+(u)=2A_0^2[\Theta(u+a)-\Theta(u-a)]. \label{eq:pluspulse}
\end{equation}

 \begin{figure}[H]
      \centering
      \includegraphics[scale=0.5]{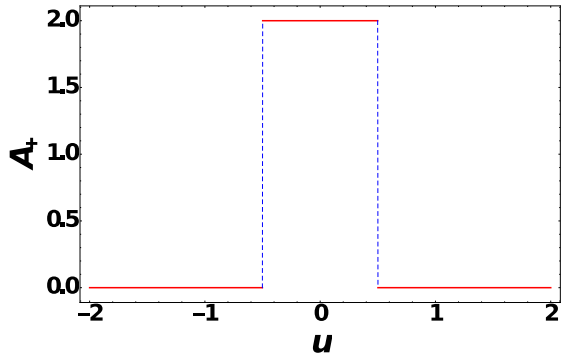}
      \caption{\small{Square pulse with $a$=0.5,$A_0$=1.}}
      \label{fig:Squarepulse}
  \end{figure}

\noindent Fig.(\ref{fig:Squarepulse}) represents a sandwich wave spacetime geometry in Brinkmann coordinates. Region-I ($u\leq-a$) and Region-III ($u\geq+a$) are flat while in Region-II ($-a\leq u\leq +a$), there is a constant amplitude pulse. This is similar to Fig.(\ref{fig:BJR_sandwich_wave}) which was obtained in BJR coordinates. The wavefronts are denoted by blue dashed lines in Fig.(\ref{fig:Squarepulse}). The width and amplitude of the pulse profile is given by $2a$ and $A_0$ respectively. Comparing with Fig.(\ref{fig:BJR_sandwich_wave}) gives $u_1=-a$ and $u_2=+a$. 

 \noindent In the following subsections, we will try to illustrate how the choice of a simple profile like the square pulse lead to exactly integrable geodesic solutions which can help us in our understanding of the memory effect.

\subsection{The geodesic equations}

\noindent The geodesic equations in Brinkmann coordinates having both non-zero polarizations are given below.
\begin{align}\label{eq:combined_x}
    \ddot{x}=-\frac{1}{2}A_+(u)x-\frac{1}{2}A_\times(u)y
\end{align}
\begin{align}\label{eq:combined_y}
    \ddot{y}=\frac{1}{2}A_+(u)y-\frac{1}{2}A_\times(u)x
\end{align}
\begin{align}\label{eq:combined_v}
  \ddot{v}+\frac{1}{4}\frac{dA_+(u)}{du}(x^2-y^2)+A_+(u)(x\dot{x}-y\dot{y})+A_\times(u)(y\dot{x}+x\dot{y})+\frac{1}{2}\frac{dA_\times(u)}{du}xy=0 
\end{align}
\noindent Notice that we have used $u$ as an affine parameter. This is
  easily checked by writing down the geodesic equation for the $v$ coordinate. The overdot denotes differentiation w. r. t. $u$. The general form for $\dot{v}(u)$ can be obtained from the  geodesic Lagrangian (derived from the metric) in Eq.(\ref{eq:brinkmann}).
 \begin{equation}
     \dot{v}=\dfrac{1}{2}(\dot{x}^2+\dot{y}^2)-\dfrac{1}{4}A_+\big( x^2-y^2\big)-\dfrac{1}{2}A_\times(u)xy+\dfrac{k}{2}  \label{eq:first_integral_v}
 \end{equation}
 
 \noindent $k$ is $0$ or $1$ for null or timelike geodesics. The first integral of the coordinate $v(u)$ can be further integrated using Eqs.(\ref{eq:combined_x}) and (\ref{eq:combined_y}) to yield
  \begin{equation}
     v(u)=v_0+\frac{1}{2}(x\dot{x}+y\dot{y}+ku)  \label{eq:v_soln}
 \end{equation}

 \noindent Here, $v_0$ is the integration constant. Thus, for any pulse of a given polarization, if Eqs.(\ref{eq:combined_x}) and (\ref{eq:combined_y}) for $x(u)$ and $y(u)$ are analytically solvable, then $v(u)$ also can be analytically obtained.
 
 \noindent Corresponding to the sandwich wave geometry, we find that the R. H. S. of Eqs.(\ref{eq:combined_x}) and (\ref{eq:combined_y}) are nonzero only for Region-II of Fig.(\ref{fig:Squarepulse}). 
 
\section{Gravitational memory effect} 

\noindent Gravitational wave memory effects can be analysed using either the geodesic equations or the geodesic deviation equations or both \citep{Zhang:2017,Chak1:2020, Siddhant:2020}. The general methodology used for finding out the geodesic memory effect is enlisted below. 

\noindent \textbullet A pulse profile in chosen to qualitatively mimic a gravitational wave burst scenario. In this article, we use a square pulse as shown in Fig.(\ref{fig:Squarepulse}). 

\noindent \textbullet Setting the initial transverse velocity to zero and using the chosen pulse profile, one solves (analytically or numerically) the geodesic equations  for two or more geodesics along the spatial directions. 

\noindent \textbullet The evolution of the {\em geodesic separation} is studied. The change in the value of the geodesic separation before and after the passage of the gravitational wave pulse gives the {\em displacement memory effect}.

\noindent \textbullet Differentiating the geodesic solutions (the solutions along $x$ and $y$ should be at least $C^1$) and studying their evolution gives the {\em velocity memory effect}. In general, pulse profiles satisfying $\int_{-\infty}^{+\infty} A_+(u) du \neq 0$, always have nonzero velocity memory \citep{Flanagan:2020,Divarkala:2021}.

\noindent \textbullet Memory effect, in general, is defined by the difference in geodesic separation at infinite future ($u\to+\infty$) and infinite past ($u\to-\infty$). But, in our article, we find focusing of geodesics at finite value of $u$. The point of focusing can be increased to higher $u$-value by appropriately choosing the initial data and parameters defining the geometry of the pulse.  Such focusing in plane wave spacetimes have been studied in works like \citep{Penrose:1965,Harte:2012,Chak:2020}. This benign focusing of the geodesics, starting from an initial parallel configuration, provides a smoking gun for the presence of {\em graviational memory} in the current context.

\noindent In the following, we mainly focus our study on the geodesic equations in order to understand memory effects arising in such spacetimes. We will consider two different scenarios. First, the geodesic separation between a pair of geodesics are analyzed and then, the shape evolution of a ring of particles (each
particle moving along a geodesic) is studied. In both these cases, we will show how memory effects are encoded in the occurence of benign focusing.  Later, we discuss gravitational memory effect obtained by solving the geodesic deviation equation and show that the results are exactly similar to the geodesic analysis.

\subsection{Memory effects for plus polarization}

\noindent We begin our calculations assuming $A_\times(u)=0$. Geodesic equations for the transverse spatial coordinates $x$ and $y$ for such a profile (given in (\ref{eq:pluspulse})) become,
 \begin{gather}
  \ddot{x}+\frac{1}{2}A_+(u)x=0  \label{eq:x_sqp}\\
  \ddot{y}-\frac{1}{2}A_+(u)y=0  \label{eq:y_sqp}
  \end{gather}
  
\noindent The geodesic equations (\ref{eq:x_sqp}) and (\ref{eq:y_sqp}) resemble a   non-relativistic oscillator (or an inverted oscillator). The pulse profiles act as squared frequencies  which are {\em time} (or $u$)-dependent \citep{Chak:2020}. Solving Eqs.(\ref{eq:x_sqp}) and (\ref{eq:y_sqp}) yields,

 \begin{align}
   x(u)=%
   \begin{cases}
    \alpha & u\leq -a \\
    \alpha \cos[A_0(u+a)] & -a\leq u \leq a\\
    \alpha A_0(a-u)\sin[2aA_0]+\alpha \cos[2aA_0] & u\geq a
   \end{cases} \label{eq:xsolsqp}
\end{align}

\begin{align}
 y(u)=%
   \begin{cases}
    \beta & u\leq -a \\
    \beta \cosh[A_0(u+a)] & -a\leq u \leq a\\
    \beta A_0(u-a)\sinh[2aA_0]+\beta \cosh[2aA_0] & u\geq a
   \end{cases} \label{eq:ysolsqp}
\end{align}  

 \noindent  Eqs.(\ref{eq:xsolsqp}) and (\ref{eq:ysolsqp}) show the behaviour of a single geodesic along $x$ and $y$ directions in all the three regions. Considering two neighbouring geodesics having initial positions set to $(x_1=\alpha_1, y_1= \beta_1)$ and 
 $(x_2=\alpha_2, y_2=\beta_2)$ respectively, we find that the geodesic separation remains constant until the gravitational wave pulse arrives. After the pulse leaves, the final separation obtained (Region-III, $u>a$) is dependent on the initial position of the geodesics and also on the height and width of the pulse.  At $u=a+A_0^{-1}\cot(2aA_0)$, we note that the separation along $x$ direction [$x_2(u)-x_1(u)$] vanishes irrespective of the value of $\alpha_1$ and $\alpha_2$. No such behaviour in found along the $y$-direction. This represents focusing of geodesic trajectories at a finite value of $u$. Substituting the analytical solutions of the geodesics for the coordinates $x$ and $y$  from Eqs.(\ref{eq:xsolsqp}) and (\ref{eq:ysolsqp}) respectively in the R. H. S. of Eq.(\ref{eq:v_soln}) gives,
 \begin{equation}
 v(u)=
   \begin{cases}
    v_0+\dfrac{k}{2}u  & u\leq -a
    \vspace{0.15in}\\
   v_0+\dfrac{k}{2}u+\dfrac{A_0}{4}\bigg(\beta ^2\sinh[2(u+a)A_0]-\alpha^2\sin[2(u+a)A_0]\bigg)  & -a\leq u \leq a
   \vspace{0.15in} \\
   v_0+ \dfrac{k}{2}u -\dfrac{\alpha^2A_0}{4}\bigg(\sin[4aA_0]+2A_0(a-u)\sin^2[2aA_0]\bigg)
   \vspace{0.05in}\\
   + \dfrac{\beta^2A_0}{4}\bigg(\sinh[4aA_0]+2A_0(u-a)\sinh^2[2aA_0]\bigg)   & u\geq a
   \end{cases} \label{eq:vsolsqp}
\end{equation} 

\noindent where $k$ is equal to 0 or 1 for null or timelike geodesics respectively. Eq.(\ref{eq:vsolsqp}) shows that $v(u)$ is continuous. This is because both $x(u)$ and $y(u)$ are continuous and differentiable.  
\begin{figure}[H]
 \centering
 \begin{subfigure}[t]{0.33\textwidth}
 	\centering
 	\includegraphics[width=\textwidth]{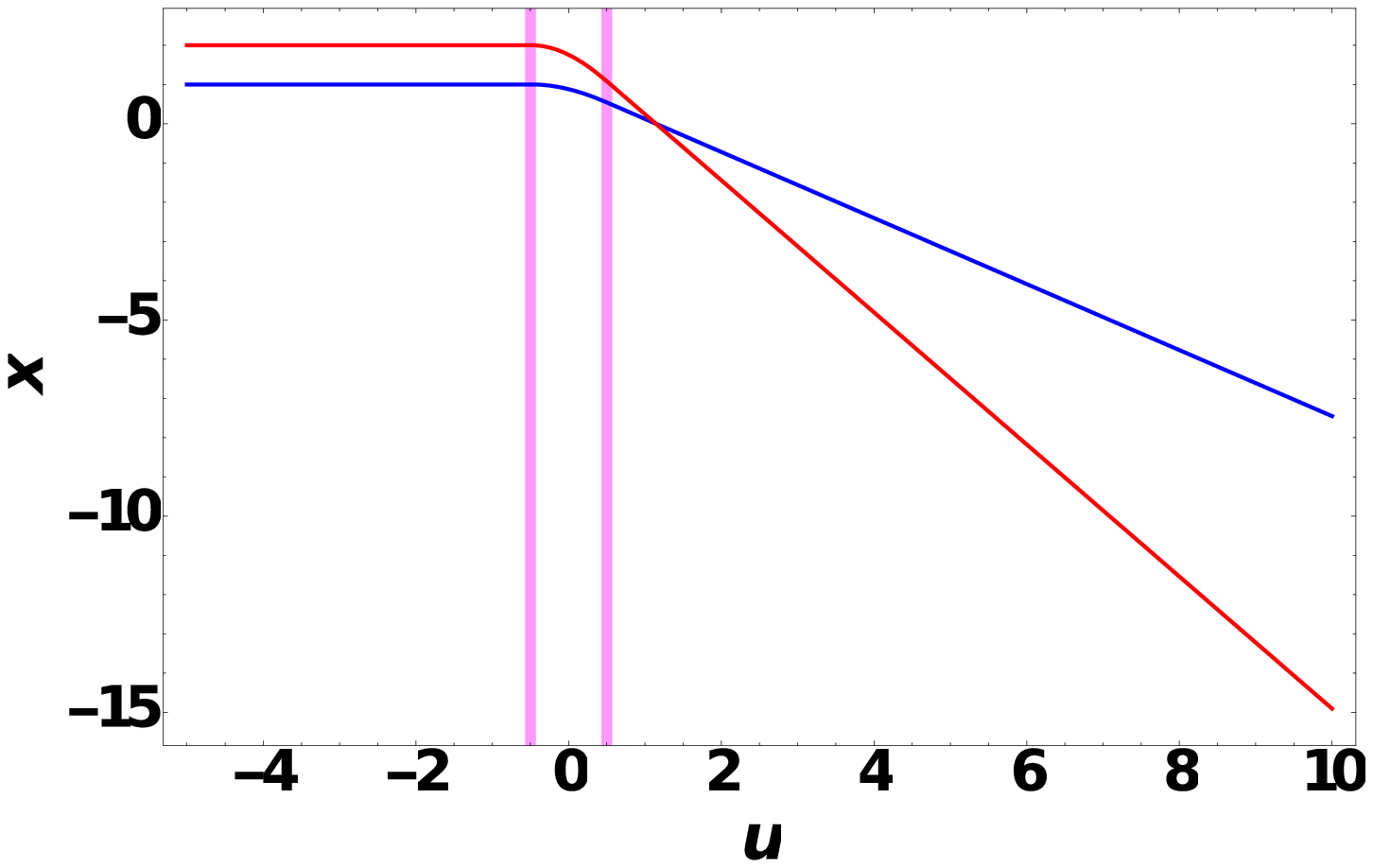}
 	\caption{ \centering{\small $x$-direction}}
 	\label{fig:x_plus}
 \end{subfigure}
 \begin{subfigure}[t]{0.33\textwidth}
 	\centering
 	\includegraphics[width=\textwidth]{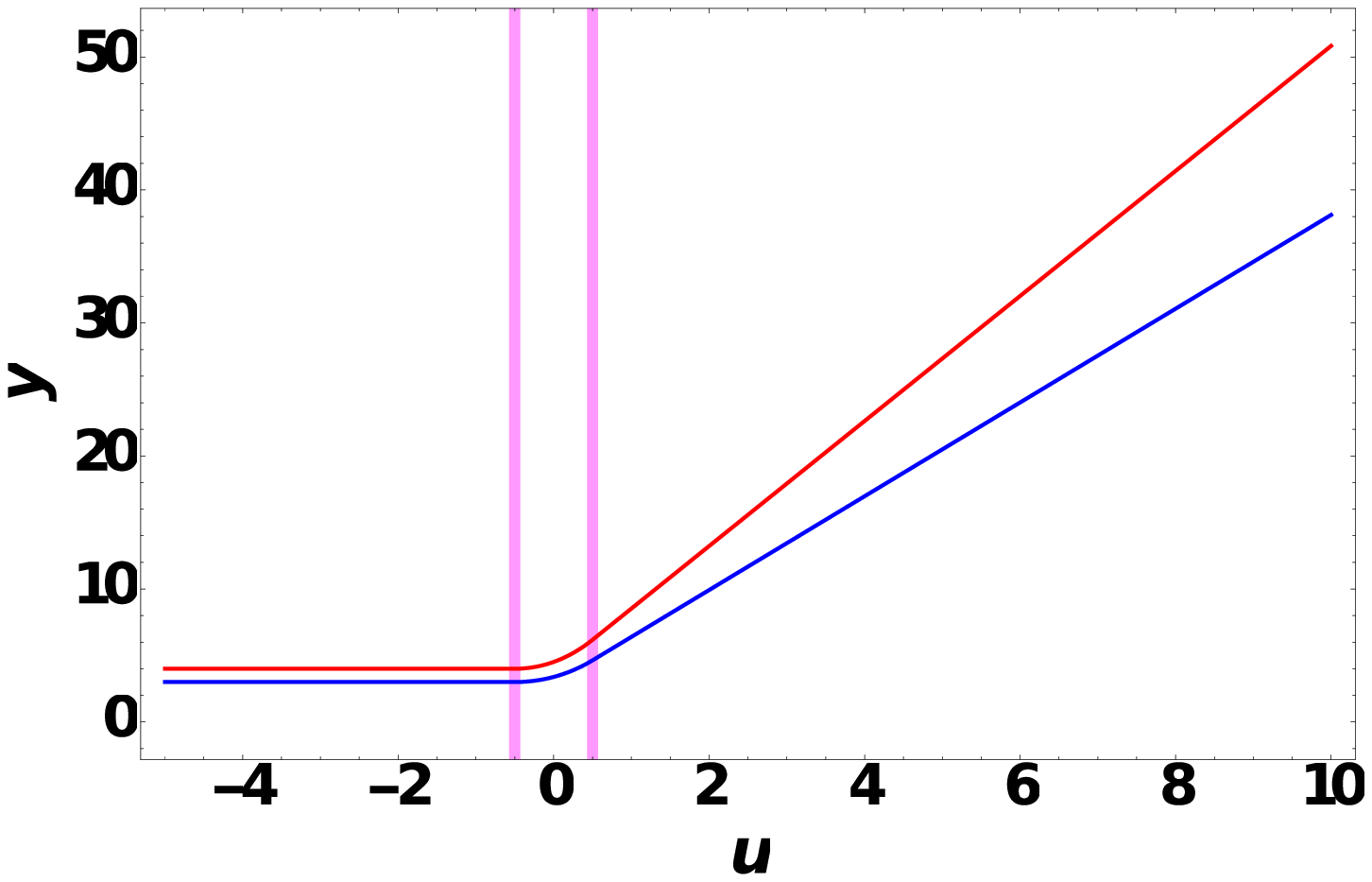}
 	\caption{\centering{\small $y$-direction}}
 	\label{fig:y_plus}
 \end{subfigure}
 \begin{subfigure}[t]{0.33\textwidth}
 	\centering
 	\includegraphics[width=\textwidth]{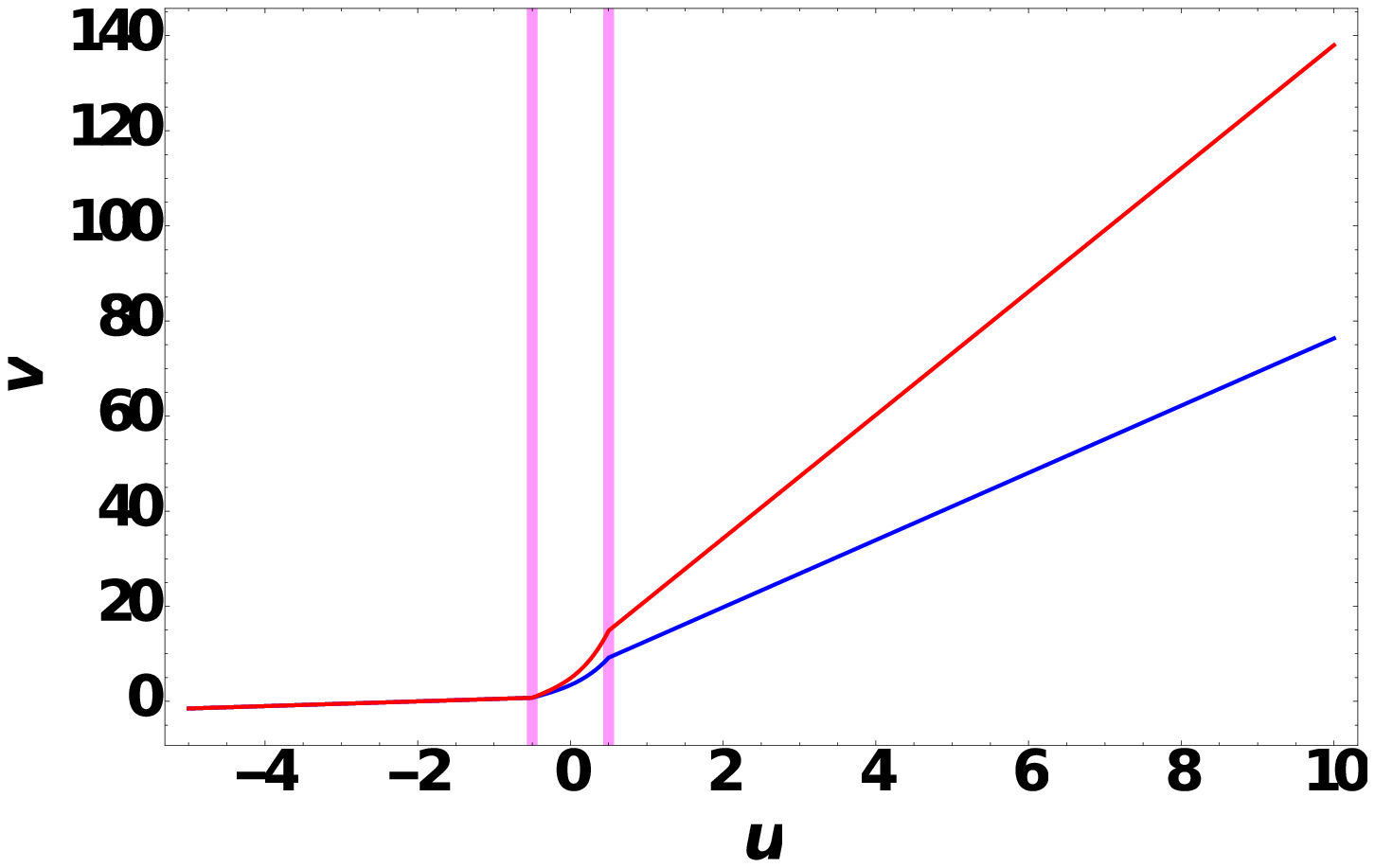}
 	\caption{\centering{\small $v$-direction}}
 	\label{fig:v_plus}
 \end{subfigure}
 \caption{\small{Displacement memory effect along $x,y,v$ directions for the first (blue) and second (red) geodesics respectively. The plots are done using the following values of the parameters: $A_0=1, a=0.5, k=1, v_0=1, \alpha= 1$(blue), 2(red), $\beta$ = 3(blue), 4(red). The curved wave region (Region-II) is shown (in all the plots) to be the
 region inside the purple vertical lines.}}
 \label{fig:Disp_memory_plus}
\end{figure}

\noindent The plots \footnote{All the plots in this article have been generated using {\em Mathematica 12}.} in Fig.(\ref{fig:Disp_memory_plus}) show the evolution of the geodesic separation along $x$, $y$ and $v$ coordinates. Along $x$ and $y$ directions we find that the separation is monotonically increasing and thus, there is non-zero {\em displacement memory}.  We also find from Fig.(\ref{fig:x_plus})
that there is a point where the trajectories meet. {As discussed earlier, this does not signify that the spacetime is singular. It only refers to the formation of a {\em benign caustic} \citep{Chak:2020}.} Along the $v$-direction shown in Fig.(\ref{fig:v_plus}), we find that the geodesics are non-differentiable at $u=-a,+a$. This is due to the discontinuous nature of the pulse profile.  

\noindent Since the geodesic solutions along $x$ and $y$ show monotonically increasing displacement memory, it is warranted that they should also exhibit velocity memory. We write down the velocities of the geodesics by differentiating Eqs.(\ref{eq:xsolsqp}), (\ref{eq:ysolsqp}) and (\ref{eq:vsolsqp}).
  \begin{align}
  \dot{x}(u)=\begin{cases} 
  0 & u\leq -a\\ 
  -\alpha A_0\sin[(u+a)A_0] & -a\leq u\leq a\\
  -\alpha A_0\sin[2aA_0] & u\geq a
  \end{cases} \label{eq:xvelsqp}
  \end{align}
  
  \begin{align}
  \dot{y}(u)=\begin{cases} 
  0 & u\leq -a\\ 
  \beta A_0\sinh[(u+a)A_0] & -a\leq u\leq a\\
  \beta A_0 \sinh[2aA_0] & u\geq a
  \end{cases} \label{eq:yvelsqp}
  \end{align}

\begin{align}
\dot{v}(u)=\begin{cases} 
  \dfrac{k}{2} & u<-a
  \vspace{0.15in}\\ 
  \dfrac{k}{2}+\dfrac{A_0^2}{2}\bigg(\beta^2\cosh[2A_0(u+a)]-\alpha^2\cos[2A_0(u+a)]\bigg) & -a< u< a
  \vspace{0.15in}\\
  \dfrac{k}{2}+\dfrac{A_0^2}{2}\bigg(\alpha^2\sin^2[2aA_0]+\beta^2\sinh^2[2aA_0]\bigg) & u> a
  \end{cases} \label{eq:vvelsqp}
\end{align}
  
\noindent Eqs.(\ref{eq:xvelsqp}) and (\ref{eq:yvelsqp}) show that both $\dot{x}(u)$ and  $\dot{y}(u)$ are continuous. But, they are not differentiable at $u=-a, +a$. Moreover, we find that $\dot{v}(u)$ is  piecewise continuous since $v(u)$ is not differentiable as shown in Fig.(\ref{fig:v_plus}).  
  
  \begin{figure}[H]
	\centering
	\begin{subfigure}[t]{0.33\textwidth}
		\centering
		\includegraphics[width=\textwidth]{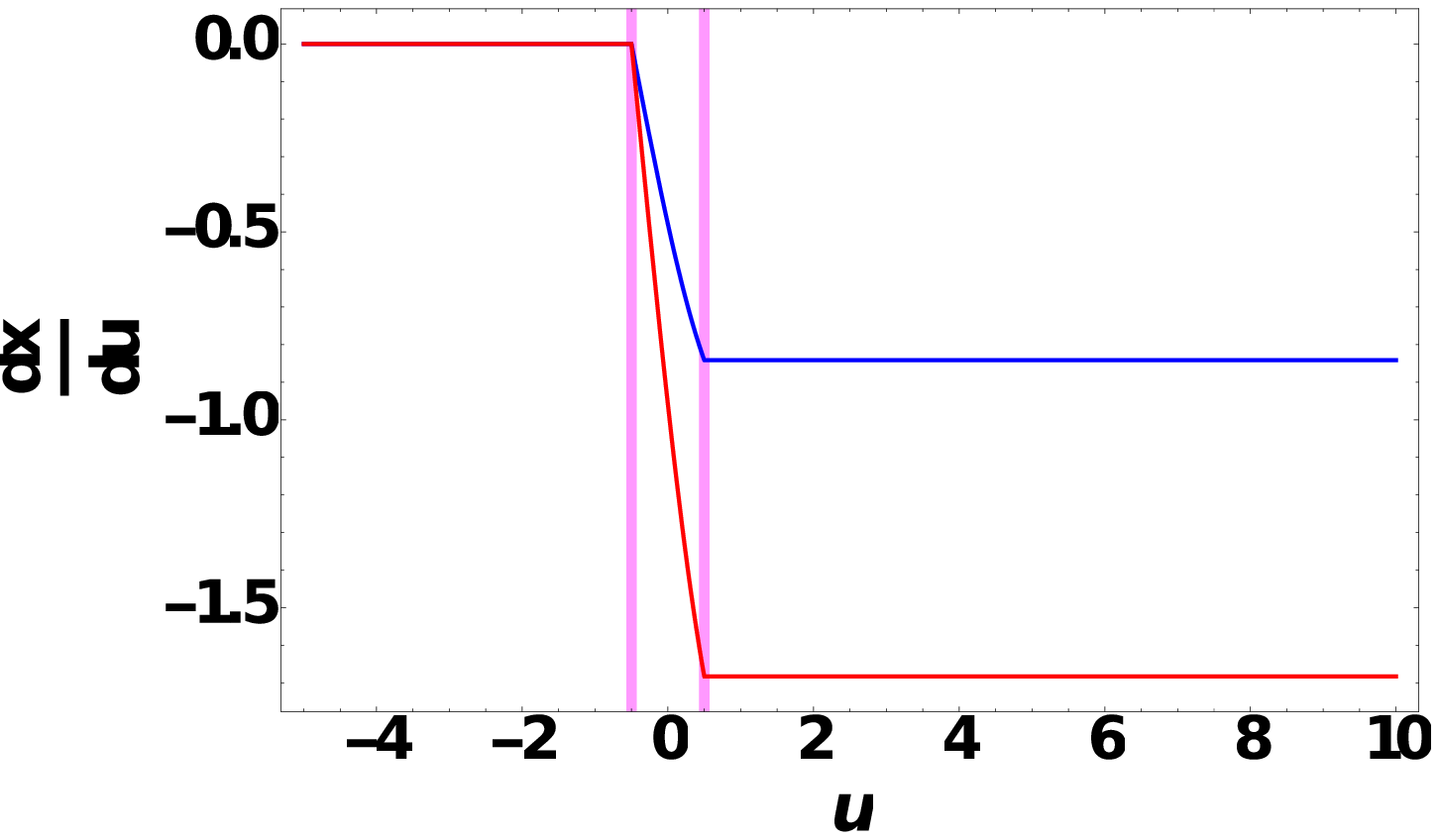}
		 \caption{\centering{\small $x$-direction}}
		\label{fig:x_vel_plus}
\end{subfigure}
	\begin{subfigure}[t]{0.33\textwidth}
		\centering
		\includegraphics[width=\textwidth]{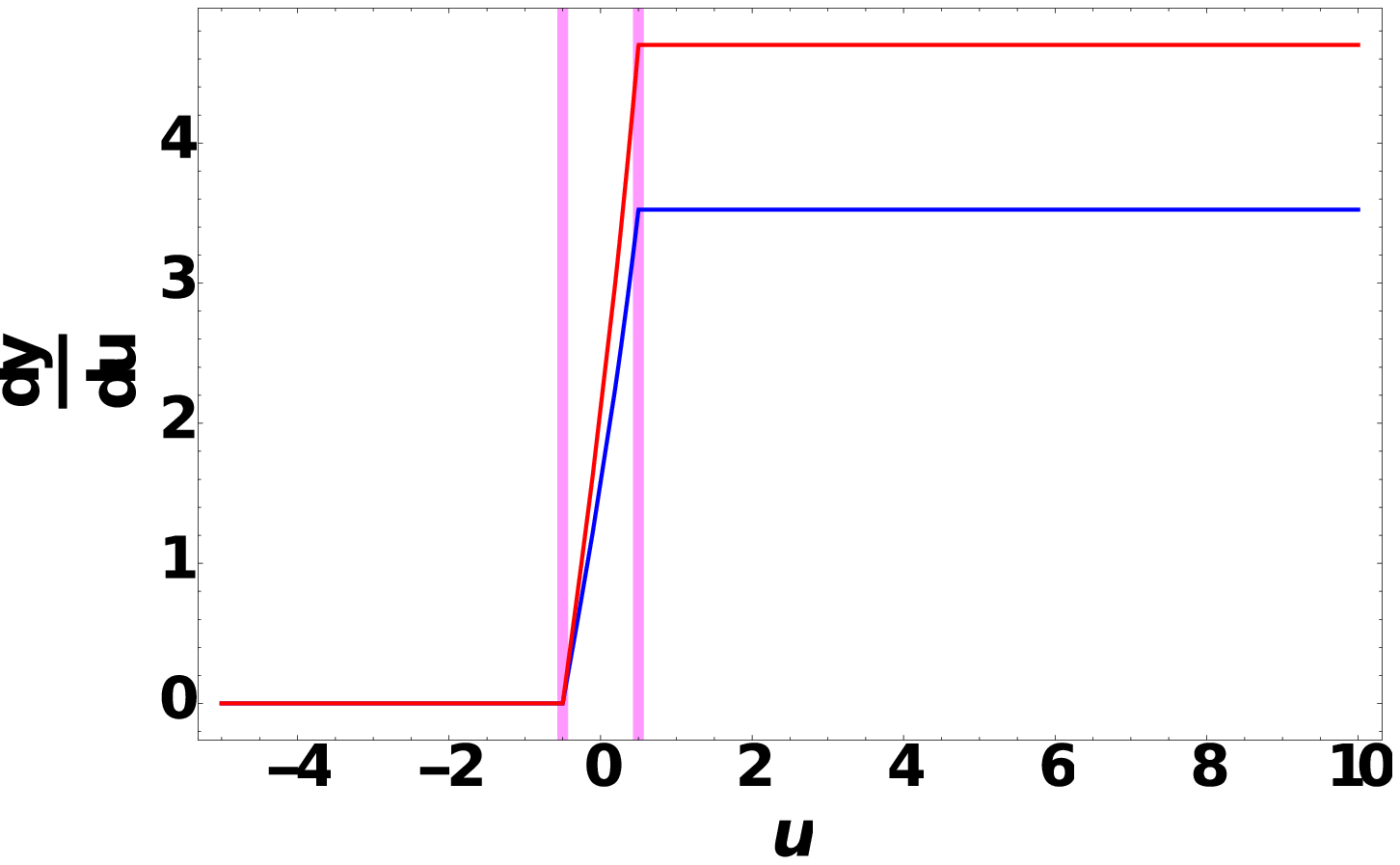}
		 \caption{\centering{\small $y$-direction}}
		\label{fig:y_vel_plus}
	\end{subfigure}
	\begin{subfigure}[t]{0.33\textwidth}
		\centering
		\includegraphics[width=\textwidth]{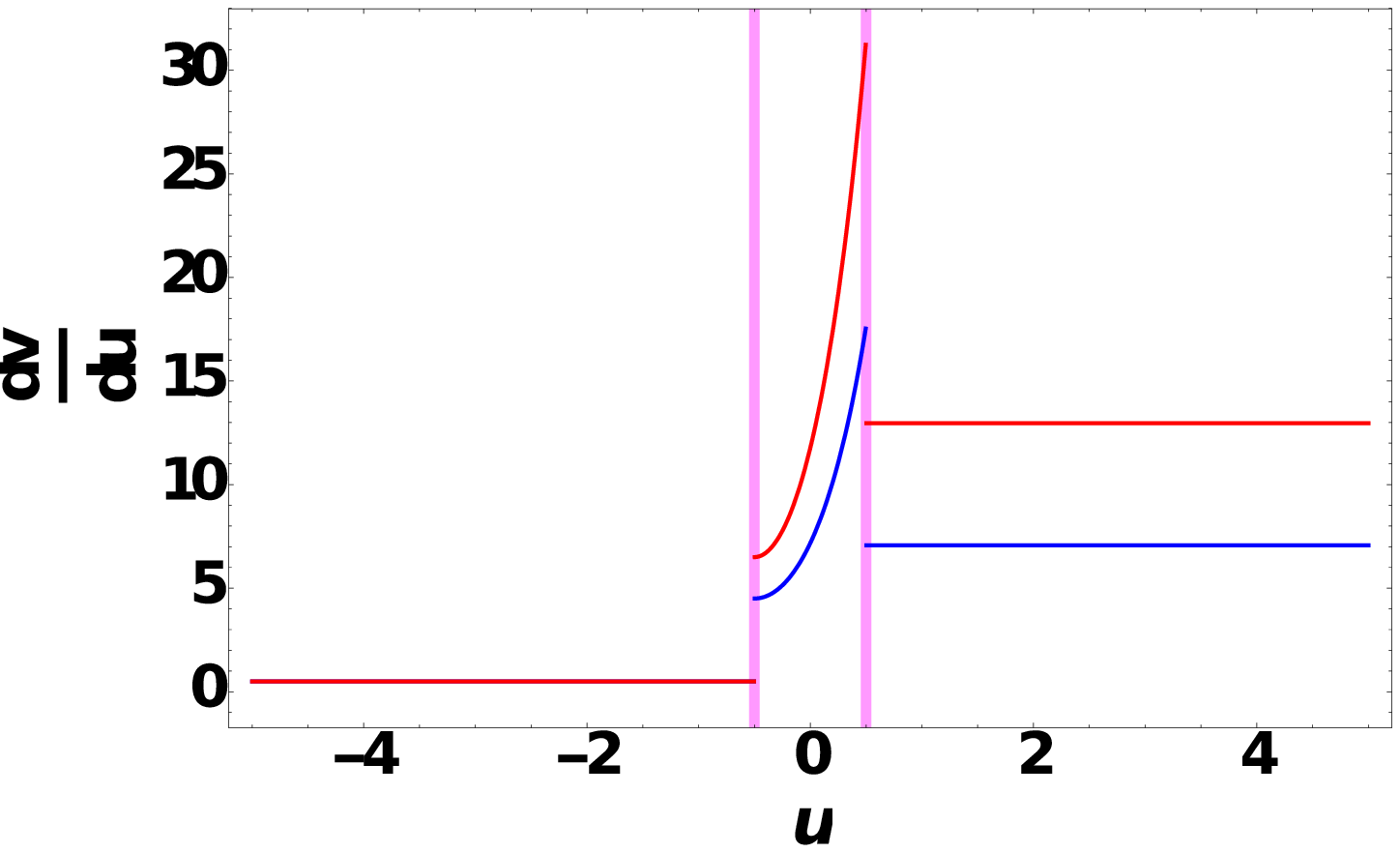}
		 \caption{\centering{\small $v$-direction}}
		\label{fig:v_vel_plus}
	\end{subfigure}
	\caption{{\small Velocity memory effect along $x,y,v$ directions.}}
	\label{fig:velocity_memory_plus}
\end{figure}
 
\noindent  Figs.(\ref{fig:velocity_memory_plus}) shows the nature of velocity profiles of the two geodesics. Along $x$ and $y$-directions, we find that initially the two geodesics are comoving. After the passage of the pulse, there is a finite velocity difference. The velocity of each geodesic builds up in the wave region and settles to a constant value (different for different geodesics). This is termed as {\em constant shift velocity memory}. As explained earlier, we observe a discontinuity at $u=-a,+a$ along the $v$-direction (Fig.(\ref{fig:v_vel_plus})).

\vspace{0.2in}

{ \noindent \textbf{Memory effect for a ring of particles?}

\noindent In our previous calculation we examined how two comoving geodesics behave in the presence of a gravitational wave pulse. Here, we try to extend this investigation for a ring of particles. From elementary gravitational physics knowledge we know that a ring of particles settles down to its initial configuration after the pulse has left. But, generally, this scenario is {\em modified} due to  the  gravitational wave memory effect. The shape of the ring is permanently distorted. In the following, we will try to analytically show how the configuration of a ring of particles changes after the departure of the pulse, using our simplified setup.

\noindent In Region-I ($u\leq -a$), the solution for $x(u)=\alpha =r\cos\phi$ and $y(u)=\beta=r\sin\phi$. Thus, the loci corresponding to the initial configuration is a circle: $x^2+y^2=r^2$. In Region-III, we have (Eqs.(\ref{eq:xsolsqp}) and (\ref{eq:ysolsqp})),
\begin{gather}
    x(u)= r [\cos (2\xi)- (\nu-1) \xi \sin(2\xi)] \cos \phi =R_1 \cos\phi \label{eq:ring_x_III}\\
    y(u) = r [\cosh (2\xi)+ (\nu-1) \xi \sinh(2\xi)] \sin \phi =R_2 \sin\phi \label{eq:ring_y_III}
\end{gather}
 
 \noindent In Eqs.(\ref{eq:ring_x_III}) and (\ref{eq:ring_y_III}), $\xi= a A_0 , u=\nu a \,(\therefore \nu>1)$. Moreover, one can check that $R_1\neq R_2$. Hence, after the pulse departs, the loci becomes an {\em elllipse}.
 \begin{equation}
     \dfrac{x^2}{R_1^2}+\dfrac{y^2}{R_2^2}=1 \label{eq:ellipse_plus}
 \end{equation} 
 
 \begin{figure}[H]
    \centering
    \includegraphics[scale=0.9]{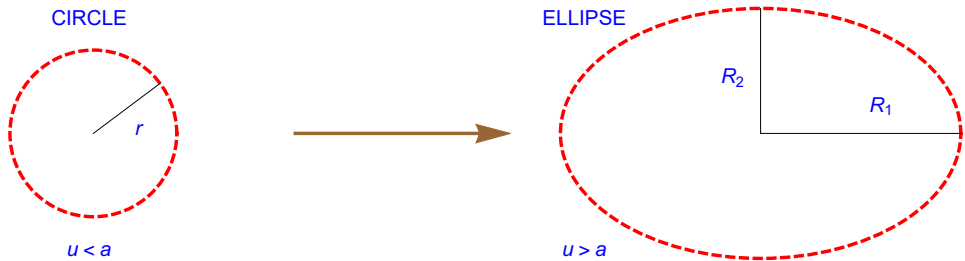}
    \caption{Change in configuration of a ring of particles from a circle to an ellipse upon the passage of a gravitational wave pulse having only plus polarization.}
    \label{fig:circle_ellipse}
\end{figure}

\noindent Fig.(\ref{fig:circle_ellipse}) demonstrates the change in the shape of the configuration induced by the pulse.  The nature of the ellipse is determined by $\xi$. Thus, the geometry of the pulse determines the change in shape of the configuration.  Now, in order to observe focusing we have to check whether $R_1$ and/or $R_2$ vanishes or not. Setting $R_1=0$ we get,
\begin{equation}
    \tan (2\xi)=\frac{1}{(\nu-1)\xi}  \label{eq:R_1}
\end{equation}
 
\noindent Similarly, setting $R_2=0$ gives,
\begin{equation}
    \tanh (2\xi) = -\frac{1}{(\nu-1)\xi} \label{eq:R_2}
\end{equation}
\noindent Eqs.(\ref{eq:R_1}) and (\ref{eq:R_2}) are transcendental and hence, analytic solutions are not possible. We try to find solutions using plots.

\begin{figure}[H]
    \centering
    \includegraphics[scale=0.7]{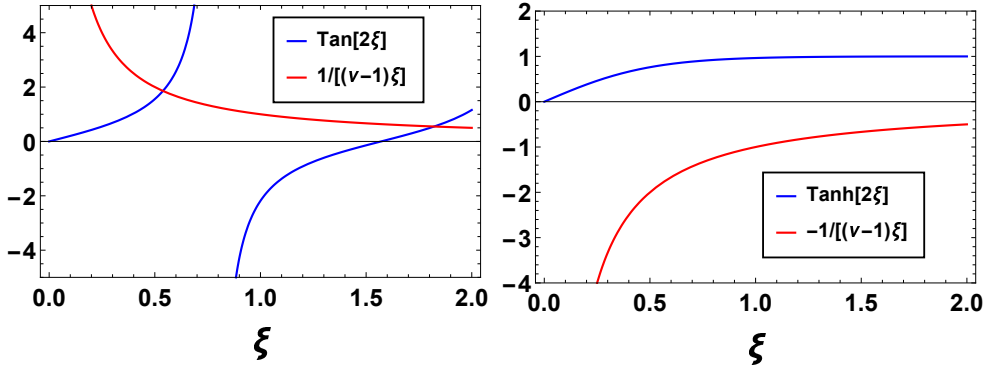}
    \caption{Solutions for the transcendental equations (\ref{eq:R_1}) and (\ref{eq:R_2}) are shown in the left ($R_1=0$) and right plots ($R_2=0$) respectively.}
    \label{fig:ring_plus}
\end{figure}
 
\noindent Fig.(\ref{fig:ring_plus}) shows that there exist solutions for $\xi$ when $R_1=0$. But, no solution exists for $R_2=0$. Thus, we find that the ellipse degenerates to a straight line along the $y$-direction. This is consistent with our previous exercise where we had shown the formation of a benign caustic as the geodesic separation along $x$-direction vanishes ({\em i.e.} $x_2-x_1=0$ in $u>a$) at a finite $u$-value. One may calculate the expansion, shear and rotation (collectively called as the {\em kinematic varianbles}) corresponding to this two-dimensional deformation.  Following the formalism given in \citep{Shaikh:2014}, the gradient of the velocity vector field can be written as,
 \begin{equation}
  \partial_j U^i= \large{ \begin{pmatrix} \frac{\theta}{2}+\sigma_+ & \sigma_\times+\omega \\  \sigma_\times - \omega & \frac{\theta}{2} - \sigma_+   \label{eq:kv_ring}
    \end{pmatrix}} 
\end{equation}
\noindent The velocity vector field $U^i=(\dot{x},\dot{y})$. Given Eq.(\ref{eq:kv_ring}), we can find all the kinematic variables using the following equations.
\begin{equation}
\begin{split}
  &  \theta = \partial_x \dot{x} +\partial_y \dot{y} \hspace{2.8cm} \sigma_+= \frac{1}{2}(\partial_x \dot{x} -\partial_y \dot{y}) \\
  & \sigma_\times= \frac{1}{2}(\partial_y \dot{x} +\partial_x \dot{y}) \hspace{2cm} \omega=\frac{1}{2}(\partial_y \dot{x} -\partial_x \dot{y})
\end{split}
\end{equation}

\noindent From Eqs.(\ref{eq:ring_x_III}) and (\ref{eq:ring_y_III}) we find that
\begin{equation}
  \theta= \bigg(\dfrac{\dot{R}_1}{R_1}+\dfrac{\dot{R}_2}{R_2}\bigg) \hspace{1.5cm} \sigma_+=\dfrac{1}{2}\bigg(\dfrac{\dot{R}_1}{R_1}-\dfrac{\dot{R}_2}{R_2}\bigg)  \hspace{1.5cm} \sigma_\times=\omega=0. \label{eq:kv_plus}
\end{equation}

\noindent Setting $R_1=0$, we find that $\theta\to-\infty, \sigma_+\to-\infty$. This is shear-induced focusing which was earlier shown in the context of ${\cal B}$-memory by the present authors in \citep{Chak:2020}. Since the expansion variable captures the area of the deformation \cite{Kar:2006}, its value diverging to minus infinity (focusing) is in agreement with the vanishing of the area of the ellipse (straight line along the $y$-axis of length $R_2$) when $R_1= 0$.

}
 
 \subsection{Memory effects for cross polarization}

\noindent We now focus our attention towards the other case of cross polarization, i.e. $A_+(u)=0, A_\times(u)\neq 0$.  The analytical form of the pulse profile is the same as was taken for the plus polarization. In this new scenario, we define normal coordinates ($X,Y$) such that, $(x+y)=X$ and $(x-y)=Y$. The geodesic equations satisfied by $X,Y$  coordinates are similar to the geodesic equations (\ref{eq:x_sqp}) and (\ref{eq:y_sqp}) for the plus polarization. Hence, the solutions for coordinates $X$ and $Y$ are similar to the ones obtained for coordinates $x$ and $y$ in the earlier case (Eqs.(\ref{eq:xsolsqp}) and (\ref{eq:ysolsqp})). Thus, the solutions in the old coordinates ($x$ and $y$) are obtained after finding solutions in the normal coordinates ($X$ and $Y$). We have,
\begin{gather}
    x(u)=%
    \begin{cases}
    \frac{1}{2}(\rho+\sigma) & u\leq -a  \vspace{0.2in}\\
    \frac{1}{2}\bigg(\rho \cos[(u+a)A_0]+\sigma \cosh[A_0(u+a)]\bigg) & -a\leq u \leq a
    \vspace{0.2in}\\
    \frac{1}{2}\bigg(\rho \cos[2aA_0]+\rho A_0(a-u)\sin[2aA_0]\\+\sigma\cosh[2aA_0]+\sigma A_0(u-a)\sinh[2aA_0]\bigg) & u\geq a
   \end{cases} \label{eq:x_cross}
\end{gather}

\vspace{0.2in}
\begin{gather}
   y(u)=%
   \begin{cases}
    \frac{1}{2}(\rho-\sigma) & u\leq -a  \vspace{0.2in}\\
    \frac{1}{2}\bigg(\rho \cos[(u+a)A_0]-\sigma \cosh[A_0(u+a)]\bigg) & -a\leq u \leq a
    \vspace{0.2in}\\
    \frac{1}{2}\bigg(\rho \cos[2aA_0]+\rho A_0(a-u)\sin[2aA_0]\\-\sigma\cosh[2aA_0]-\sigma A_0(u-a)\sinh[2aA_0]\bigg) & u\geq a
   \end{cases} \label{eq:y_cross}
\end{gather}
\noindent Eqs.(\ref{eq:x_cross}) and (\ref{eq:y_cross}) reveal that the focusing $u$-value, unlike the earlier case,  will also depend on the initial positions of the geodesics. This is quantified in the following equations. 
\begin{gather}
    u_{x}=a+A_0^{-1}\bigg[\frac{(\rho_2-\rho_1)\cos(2aA_0)+(\sigma_2-\sigma_1)\cosh(2aA_0)}{(\rho_2-\rho_1)\sin(2aA_0)-(\sigma_2-\sigma_1)\sinh(2aA_0)}\bigg] \label{eq:u_focus_x_cross}\\
    \vspace*{0.3in}
     u_{y}=a+A_0^{-1}\bigg[\frac{(\rho_2-\rho_1)\cos(2aA_0)-(\sigma_2-\sigma_1)\cosh(2aA_0)}{(\rho_2-\rho_1)\sin(2aA_0)+(\sigma_2-\sigma_1)\sinh(2aA_0)}\bigg] \label{eq:u_focus_y_cross}
\end{gather}

\noindent $\rho_1,\sigma_1$, $\rho_2,\sigma_2$ are initial positions of a pair of geodesics in normal coordinates along $X$ and $Y$ directions respectively. {Eqs.(\ref{eq:u_focus_x_cross}) and (\ref{eq:u_focus_y_cross}) show that the focusing occurs in Region-III depending on the initial values of the geodesics and shape of the pulse.} The solution for the $v$ coordinate is provided below.

\begin{gather}
   v(u)=%
   \begin{cases}
    v_0+\dfrac{k}{2}u & u\leq -a  \vspace{0.2in}\\
    v_0+\dfrac{k}{2}u+\dfrac{A_0}{8}\bigg(-\rho^2\sin[2(u+a)A_0]+\sigma^2\sinh[2(u+a)A_0]\bigg) & -a\leq u \leq a
    \vspace{0.2in}\\
    v_0+\dfrac{k}{2}u+\dfrac{A_0}{8}\bigg(-\rho^2 \big(\sin[4aA_0]+2A_0(a-u)\sin^2[2aA_0]\big)\\
    +\sigma^2\big(\sinh[4aA_0]+2A_0(u-a)\sinh^2[2aA_0]\big)\bigg) & u\geq a
   \end{cases} \label{eq:v_cross}
\end{gather}
 \begin{figure}[H]
	\centering
	\begin{subfigure}[t]{0.32\textwidth}
		\centering
		\includegraphics[width=\textwidth]{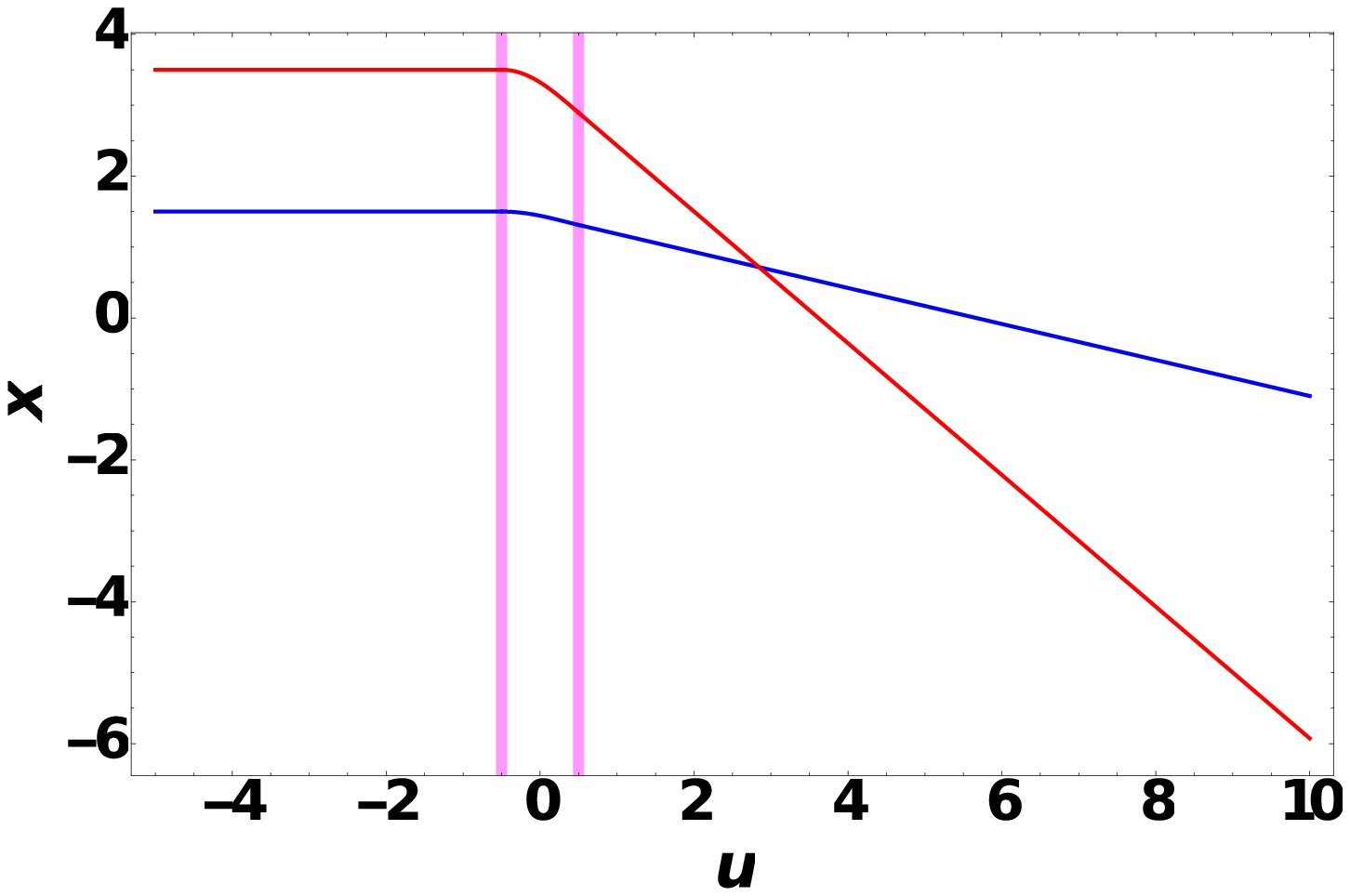}
		 \caption{\centering{ \small $x$-direction}}
		\label{fig:x_cross_geo}
\end{subfigure}\hspace{0.3em}
	\begin{subfigure}[t]{0.32\textwidth}
		\centering
		\includegraphics[width=\textwidth]{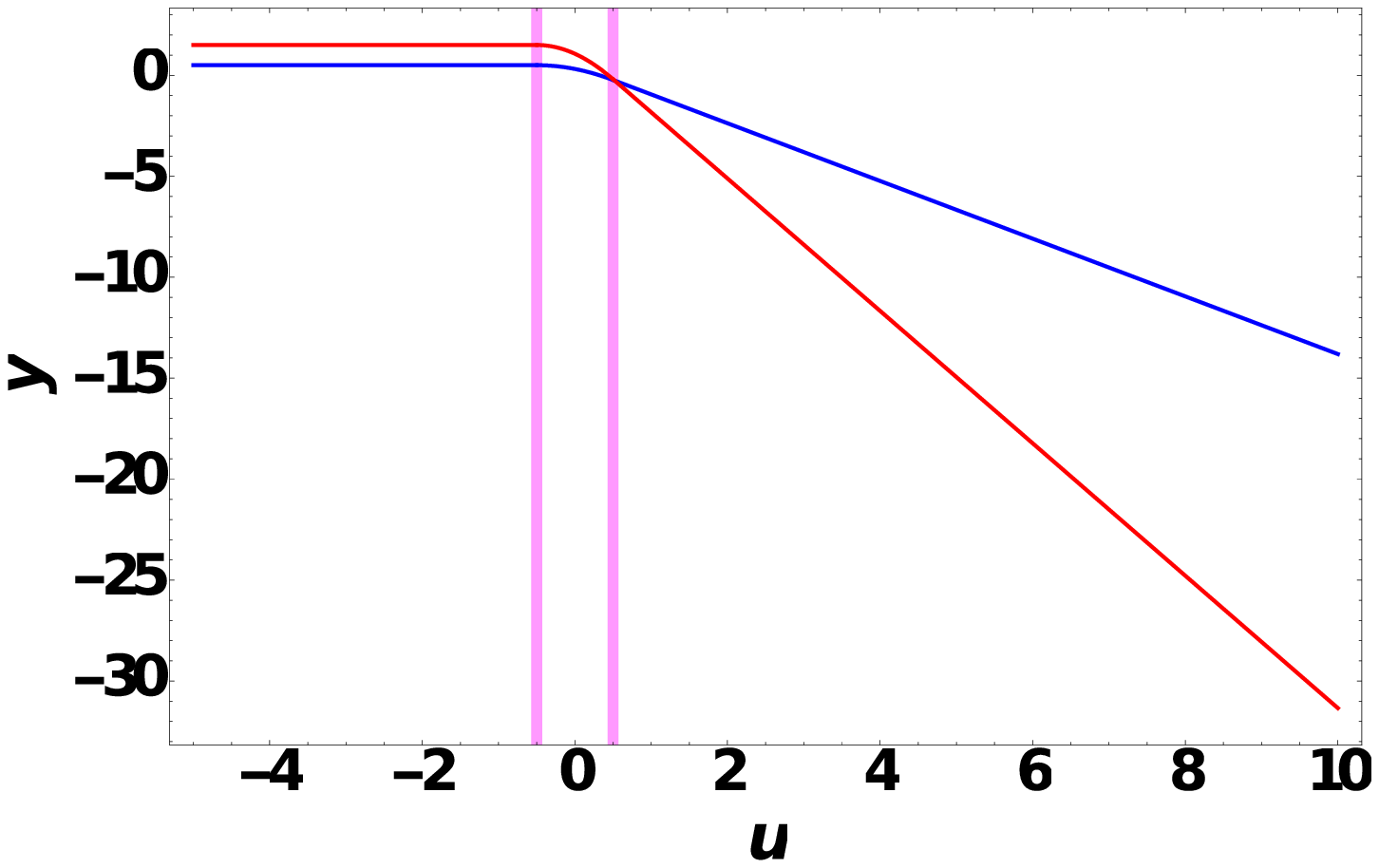}
		 \caption{\centering{ \small $y$-direction}}
		\label{fig:y_cross_geo}
	\end{subfigure}
	\begin{subfigure}[t]{0.32\textwidth}
		\centering
		\includegraphics[width=\textwidth]{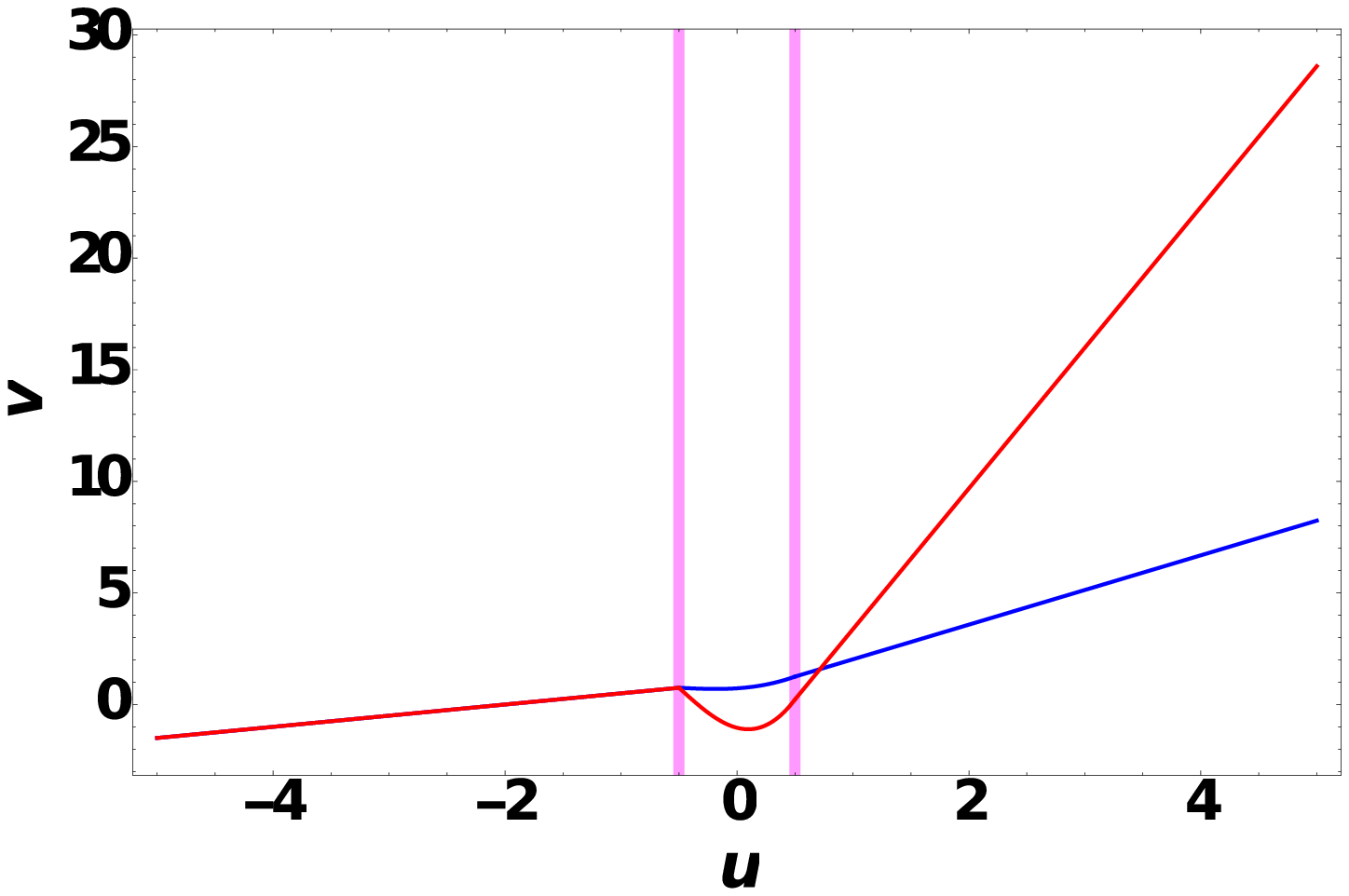}
		 \caption{\centering{\small $v$-direction}}
		\label{fig:v_cross_geo}
	\end{subfigure}
	\caption{\small{Displacement memory effect along $x,y,v$ directions for the first (blue) and second (red) geodesics respectively. The plots are done using the following values of the parameters: $A_0=1, a=0.5, k=1, v_0=1, \rho= 2$(blue), 5(red), $\sigma$ = 1(blue), 2(red).}}
	\label{fig:cross_sqp}
\end{figure}

\noindent Figs.(\ref{fig:x_cross_geo}) and (\ref{fig:y_cross_geo}) show monotonic displacement memory along $x$ and $y$ directions respectively. The two trajectories meet at different values of $u$ since they depend on the values of initial positions of each geodesic. In Fig.(\ref{fig:v_cross_geo}), we find that the nature is continuous but not differentiable at the edges of the pulse ($u=-a, +a$). The expressions for the velocities are given below.
\begin{gather}
   \dot{x}(u)=%
   \begin{cases}
    0 & u\leq -a \vspace{0.1in}\\
    \dfrac{A_0}{2}\bigg(-\rho \sin[(u+a)A_0]+\sigma \sinh[(u+a)A_0]\bigg) & -a\leq u \leq a \vspace{0.1in}\\
    \dfrac{A_0}{2}\bigg(-\rho \sin[2aA_0]+\sigma \sinh[2aA_0]\bigg) & u\geq a
   \end{cases} \label{eq:x_vel_cross}
\end{gather}

\begin{gather}
   \dot{y}(u)=%
   \begin{cases}
    0 & u\leq -a \vspace{0.1in}\\
    -\dfrac{A_0}{2}\bigg(\rho \sin[(u+a)A_0]+\sigma \sinh[(u+a)A_0]\bigg) & -a\leq u \leq a \vspace{0.1in}\\
    -\dfrac{A_0}{2}\bigg(\rho \sin[2aA_0]+\sigma \sinh[2aA_0]\bigg) & u\geq a
   \end{cases} \label{eq:y_vel_cross}
\end{gather}

\begin{gather}
   \dot{v}(u)=%
   \begin{cases}
    \dfrac{k}{2} & u<-a \vspace{0.15in}\\
    \dfrac{k}{2}+\dfrac{A_0^2}{4}\bigg(-\rho^2\cos[2(u+a)A_0]+\sigma^2\cosh[2(u+a)A_0]\bigg) & -a<u<a \vspace{0.15in}\\
    \dfrac{k}{2}+\dfrac{A_0^2}{4}\Big(\rho^2\sin^2[2aA_0]+\sigma^2\sinh^2[2aA_0]\Big) & u>a
   \end{cases} \label{eq:v_vel_cross}
\end{gather}

\noindent Similar to the previous analysis for the plus polarization, we find here that the velocities $\dot{x}(u)$ and $\dot{y}(u)$ are continuous but not differentiable. The strict inequality in Eq.(\ref{eq:v_vel_cross}) for $\dot{v}(u)$  follows from Eq.(\ref{eq:first_integral_v}).
\begin{figure}[H]
	\centering
	\begin{subfigure}[t]{0.32\textwidth}
		\centering
		\includegraphics[width=\textwidth]{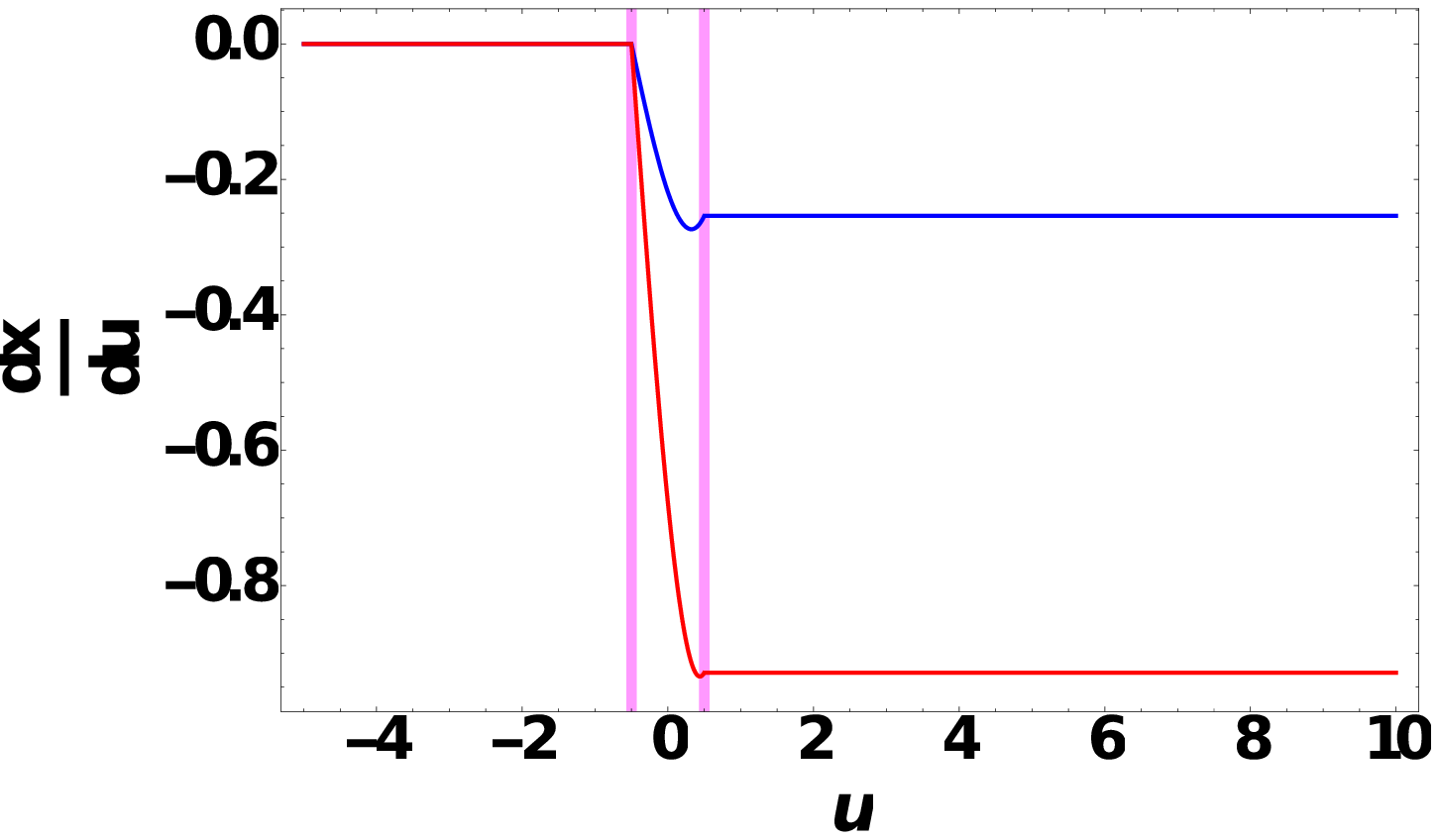}
		 \caption{\centering{\small $x$-direction}}
		\label{fig:x_vel_cross}
\end{subfigure}
	\begin{subfigure}[t]{0.32\textwidth}
		\centering
		\includegraphics[width=\textwidth]{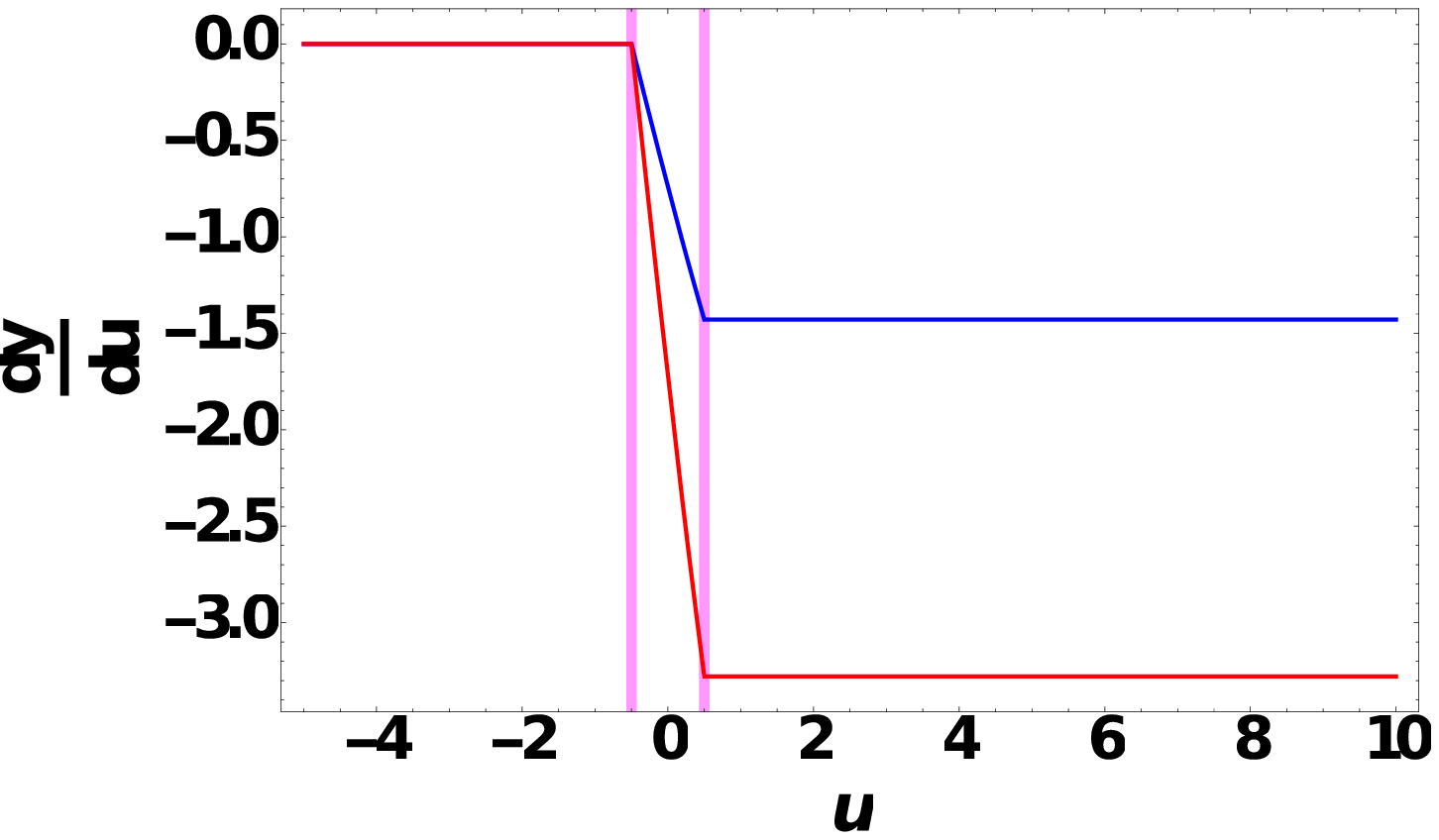}
		 \caption{\centering{\small $y$-direction}}
		\label{fig:y_vel_cross}
	\end{subfigure}
	\begin{subfigure}[t]{0.32\textwidth}
		\centering
		\includegraphics[width=\textwidth]{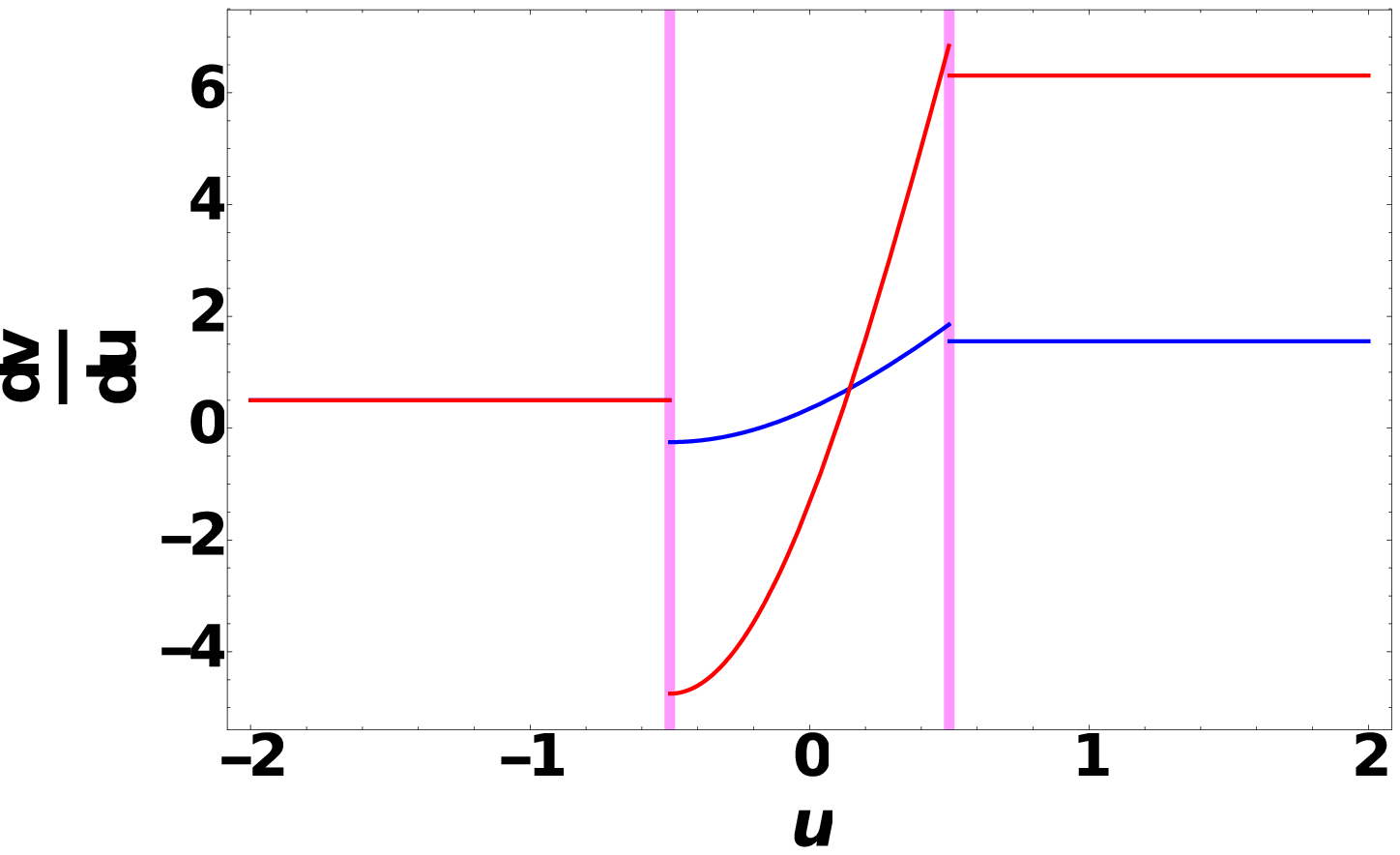}
		 \caption{\centering{\small $v$-direction}}
		\label{fig:v_vel_cross}
	\end{subfigure}
	\caption{\centering{{\small Velocity memory effect along $x,y,v$ directions}}}
	\label{fig:vel_cross_sqp}
\end{figure}

\noindent Figs.(\ref{fig:x_vel_cross}) and (\ref{fig:y_vel_cross}) show similar behaviour as noted earlier in Figs.(\ref{fig:x_vel_plus}) and (\ref{fig:y_vel_plus}) for plus polarization. There is constant shift velocity memory  along both these directions. For the $v$-direction shown in Fig.(\ref{fig:v_vel_cross}), we find that the velocities are not continuous at $u=-a,+a$ due to the analytical form of $A_\times(u)$.

\begin{figure}[H]
    \centering
    \includegraphics[scale=0.9]{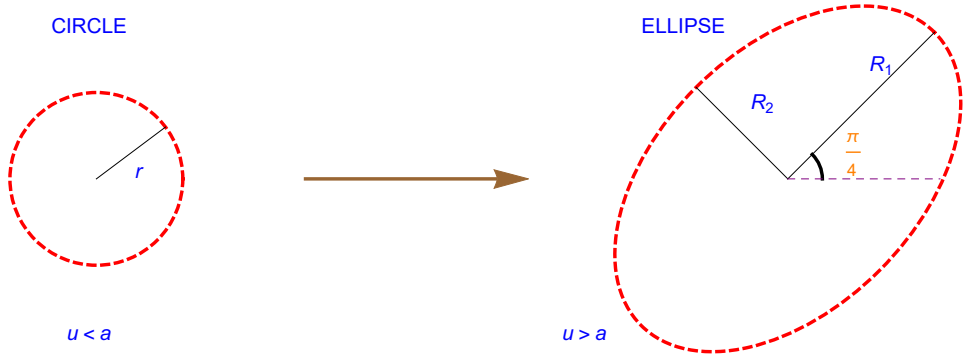}
    \caption{Change in configuration of a ring of particles from a circle to an ellipse upon the passage of a gravitational wave pulse having only cross polarization.}
    \label{fig:circle_ellipse_cross}
\end{figure}

{
\noindent For a ring of particles, we follow the same procedure as was done for the plus polarization. We first perform the calculation in normal ($X,Y$) coordinates and then revert back to the old $(x,y)$ coordinates. Before the pulse has arrived ($u\leq-a$), we have
\begin{gather}
    X=\rho= r\cos\theta \hspace{2cm} Y= \sigma=r\sin\theta
\end{gather}
\noindent Thus, the loci is a circle: $X^2+Y^2=r^2$. Transforming to old coordinates we find that the loci is again a circle with a different radius, $x^2+y^2=\bigg(\dfrac{r}{\sqrt{2}}\bigg)^2$. 

\noindent In Region-III ($u\geq a$), we get the loci of an ellipse using normal coordinates having axes lengths $2R_1$ and $2R_2$ (look at Eq.(\ref{eq:ellipse_plus})). Reverting back to the old coordinates we find,
\begin{equation}
    \dfrac{\bigg(\dfrac{x}{\sqrt{2}}+\dfrac{y}{\sqrt{2}}\bigg)^2}{\bigg(\dfrac{R_1}{\sqrt{2}}\bigg)^2}+\dfrac{\bigg(\dfrac{x}{\sqrt{2}}-\dfrac{y}{\sqrt{2}}\bigg)^2}{\bigg(\dfrac{R_2}{\sqrt{2}}\bigg)^2}=1 \label{eq:ellipse_cross}
\end{equation}

\vspace{0.1in}

\noindent The loci given by Eq.(\ref{eq:ellipse_cross}) also corresponds to an ellipse having center at $\{x,y\}=\{0,0\}$ with axes lengths $\sqrt{2}R_1$ and $\sqrt{2}R_2$. It is rotated by an angle of $\pi/4$ w. r. t. the $x$-axis along one of the axes of the ellipse having length $\sqrt{2}R_1$ (Fig.(\ref{fig:circle_ellipse_cross})). Similar to the earlier analysis, $R_1=0$ is a possible scenario. But now, the straight line is parallel to $R_2$ which is rotated by an angle of $3\pi/4$ w. r. t. the $x$-axis. 
Calculating the kinematic variables one finds that the expansion is the same as that for the plus polarization. The shear corresponding to this deformation is now given by $\sigma_\times\neq0, \sigma_+=0$. The functional form of $\sigma_\times$ is similar to the expression given in Eq.(\ref{eq:kv_plus}). This can also be understood intuitively from Eq.(\ref{eq:ellipse_cross}) and Fig.(\ref{fig:circle_ellipse_cross}) as the ellipse now has been rotated w. r. t. the $x$-direction. }

{
\subsection{Memory effects for both plus and cross polarizations}

\noindent Finally, let us consider the scenario when both polarizations $A_+$ and $A_\times$ are present and also have the same pulse profile.  
 If there is a phase shift by $\pi/2$ between the two polarizations, the plane gravitational wave is said to be circularly polarized \citep{Zhang:2018vel}. But, here,  we consider $A_+=A_\times\neq0$ (no phase difference) which correspond to {\em linear polarization} \citep{Schutz:1987}.  
The resultant geodesic equations along $x$ and $y$ directions in this scenario are already given previously in Eqs.(\ref{eq:combined_x}) and (\ref{eq:combined_y}). Geodesic equations in Region-II (where the pulse is present) is given below.
\begin{equation}
    \ddot{x}= -A_0^2 (x+y)\hspace{2cm}  \ddot{y} =A_0^2 (y-x) \label{eq:geo_both}
\end{equation}

\noindent Taking double derivative on both sides of the equations in (\ref{eq:geo_both}) we get,
\begin{equation}
\ddddot{x}=2A_0^4 x \hspace{2cm}  \ddddot{y} =2A_0^4 y \label{eq:geo_II_both}
\end{equation}

\noindent In Region-I ($u\leq-a$), the solution is $x(u)=\epsilon, y(u)=\delta$. In the presence of the pulse (Region-II, $-a\leq u\leq +a$), the geodesic solution is,
\begin{gather}
    x(u)= C_1 \cosh[m u A_0] +C_2 \sinh [m u A_0] +C_3 \cos [m u A_0] +C_4 \sin[m u A_0] \label{eq:x_both_pol}\\
    y(u) =  C_5 \cosh[m u A_0] +C_6 \sinh [m u A_0] +C_7 \cos [m u A_0] +C_8 \sin[m u A_0] \label{eq:y_both_pol}
\end{gather}

\noindent  In Eqs.(\ref{eq:x_both_pol}) and (\ref{eq:y_both_pol}), we have $m=2^{1/4}$. The constants of integration ($C_1-C_8$) are given below.

\begin{equation}
    \begin{split}
      &  C_1= \frac{[\epsilon ( m^2-1)-\delta
   ]\cosh [m a A_0 ]}{2 m^2}  \hspace{1.5cm} 
   C_2= \frac{[\epsilon ( m^2-1)-\delta
   ] \sinh [m a A_0 ]}{2 m^2}\\
& C_3=\frac{[\delta +(m^2+1) \epsilon ]
   \cos [m a A_0 ]}{2 m^2} \hspace{1.7cm} 
   C_4= -\frac{[\delta +(m^2+1) \epsilon ]
   \sin [m a A_0 ]}{2 m^2} \\
      &  C_5= \frac{[\delta (m^2+1)-\epsilon] \cosh [m a A_0 ]}{2
   m^2}   \hspace{1.5cm} 
   C_6= \frac{[\delta (m^2+1)-\epsilon]  \sinh [m a A_0 ]}{2
   m^2} \\
& C_7= \frac{[\delta (m^2-1)+\epsilon]  \cos [m a A_0 ]}{2 m^2} \hspace{1.75cm} 
   C_8= -\frac{[\delta (m^2-1)+\epsilon]\sin [m a A_0 ]}{2 m^2} \label{eq:constant_x,y}
    \end{split} 
\end{equation}

\noindent The above expressions in Eq.(\ref{eq:constant_x,y}) are determined by comparing the coefficients in Eq.(\ref{eq:geo_both}) and also using the boundary conditions at $u=-a$ given as:
$$x(-a)=\epsilon, \dot{x}(-a)=0, y(-a)=\delta,\dot{y}(-a)=0.$$

\noindent Finally, the solution in Region-III is,
\begin{equation}
        x(u)= C_9 u+C_{10} \hspace{3cm} y(u)= C_{11}u +C_{12} \label{eq:geo_soln_both_III}
\end{equation}
\noindent The constants ($C_9-C_{12}$) are obtained by assuming continuity and differentiability of the geodesic solutions at the boundary $u=+a$ of the pulse.

\begin{equation}
\begin{split}
          & C_9= \frac{A_0}{2m}\big[(\epsilon ( m^2-1)-\delta
   ) \sinh(2 m a A_0)-( \epsilon(m^2+1)+ \delta)\sin(2 m a A_0)\big] \\
    & C_{10}= \frac{(\epsilon(m^2-1)-\delta)}{2m^2}\big[\cosh (2maA_0)-maA_0\sinh(2maA_0)\big]\\
 & \hspace{1cm}  +\frac{(\epsilon(m^2+1)+\delta)}{2m^2}\big[\cos (2maA_0)+maA_0\sin(2maA_0)\big]\\
      & C_{11}=\frac{A_0}{2m}\big[(\delta(m^2+1)-\epsilon) \sinh(2 m a A_0)-(\delta(m^2-1) +\epsilon
   )\sin(2 m a A_0)\big] \\
 &   C_{12}= \frac{(\delta(m^2+1)-\epsilon)}{2m^2}\big[\cosh (2maA_0)-maA_0\sinh(2maA_0)\big]\\
 & \hspace{1cm}  +\frac{(\delta(m^2-1)+\epsilon)}{2m^2}\big[\cos (2maA_0)+maA_0\sin(2maA_0)\big]
\end{split}
\end{equation}

\noindent We, now, try to understand the nature of memory effect exhibited by the geodesic solutions in this scenario.

\begin{figure}[H]
	\centering
	\begin{subfigure}[t]{0.45\textwidth}
		\centering
		\includegraphics[width=\textwidth]{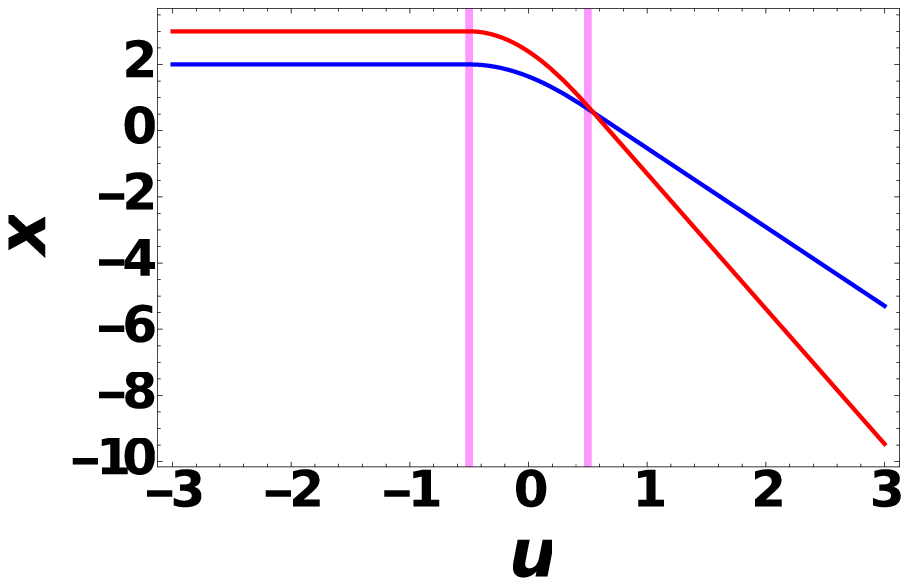}
		 \caption{\centering{\small $x$-direction}}
		\label{fig:x_both}
\end{subfigure} \hspace{1cm}
	\begin{subfigure}[t]{0.45\textwidth}
		\centering
		\includegraphics[width=\textwidth]{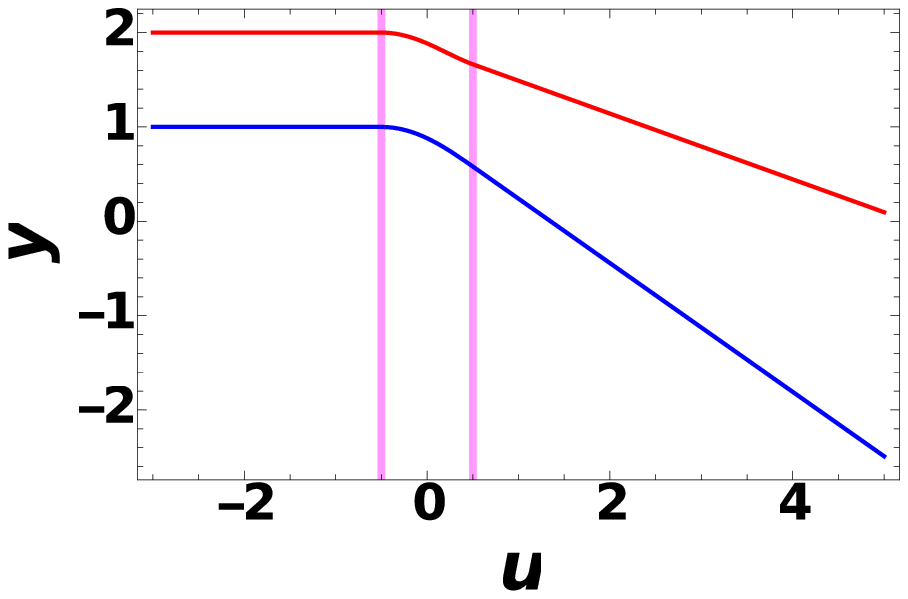}
		 \caption{\centering{\small $y$-direction}}
		\label{fig:y_both}
	\end{subfigure}
	\caption{{\small Memory effect along $x,y$ directions having both $A_+$ and $A_\times$ polarizations. The plots are done using the following values of the parameters: $A_0=1, a=0.5, \epsilon=  2$(blue),3(red), $\delta$ = 1(blue), 2(red). The constants of integration are: $C_1= -0.0717032$(blue),$-0.316513$(red), $C_2= -0.038232$(blue), $-0.168764$(red), $C_3 =1.70699$(blue), $2.70692$(red), $C_4= -1.15434$(blue), $-1.83054$(red), $C_5=0.173107$ (blue), $0.76413$(red), $C_6=0.0923002$(blue),$0.407433$(red), $C_7= 0.707059$(blue), $1.12124$(red), $C_8= -0.478144$(blue), $-0.758234$(red), $C_9= -2.38178$(blue), $-4.08101$(red), $C_{10}= 1.84942$(blue), $2.77691$(red), $C_{11}= -0.682551$(blue), $-0.348423$(red), $C_{12}= 0.921929$(blue), $1.8383$(red). }}
	\label{fig:vel_cross_sqp}
\end{figure}

\noindent The plots in Fig.(\ref{fig:x_both}) and (\ref{fig:y_both}) show monotonically increasing displacement memory effect. One can also obtain the plot for $v$-coordinate using Eq.(\ref{eq:v_soln}). The velocity plots for the solutions are shown in Fig.(\ref{fig:x_both_vel}) and (\ref{fig:y_both_vel}).

\begin{figure}[H]
	\centering
	\begin{subfigure}[t]{0.45\textwidth}
		\centering
		\includegraphics[width=\textwidth]{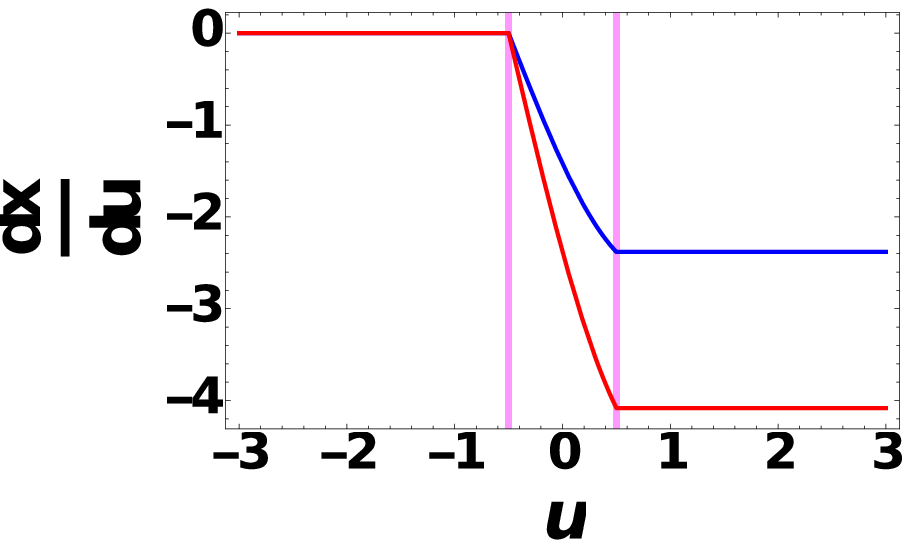}
		 \caption{\centering{\small $x$-direction}}
		\label{fig:x_both_vel}
\end{subfigure} \hspace{1cm}
	\begin{subfigure}[t]{0.45\textwidth}
		\centering
		\includegraphics[width=\textwidth]{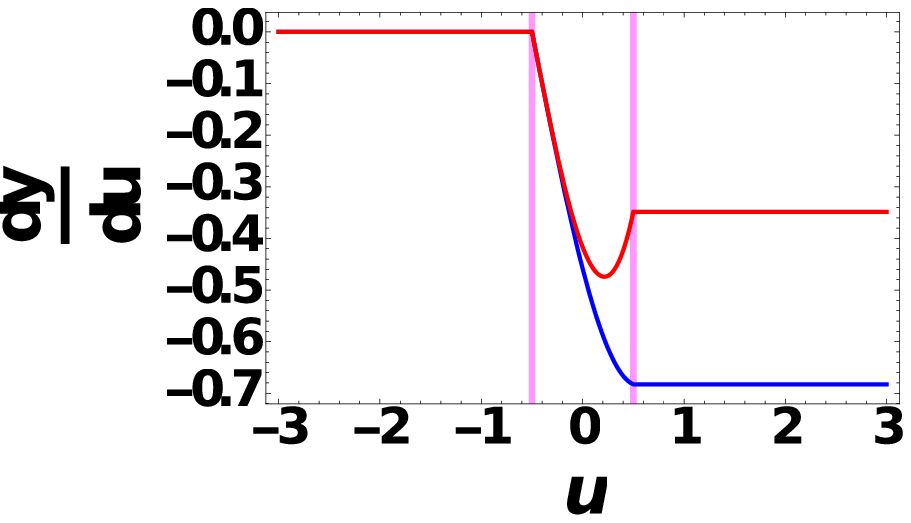}
		 \caption{\centering{\small $y$-direction}}
		\label{fig:y_both_vel}
	\end{subfigure}
	\caption{\centering{{\small Velocity memory effect along $x,y$ directions having both $A_+$ and $A_\times$ polarizations.}}}
	\label{fig:vel_cross_sqp}
\end{figure}

\noindent Similar to previous scenarios, we again find constant shift velocity memory effects. In case of a ring of particles, one can check that both $\sigma_+$ and $\sigma_\times$ are nonzero. This shows that the polarization of the gravitational wave influences the final configuration.}

\subsection{Memory effects using geodesic deviation equation}
 \noindent Let us now try to analyse memory effects using the geodesic deviation equation
 which, we recall, is given by,
\begin{equation}
    (U^\alpha\nabla_\alpha) (U^\beta\nabla_\beta) \xi^\mu=-R^\mu\,_{\alpha\beta\gamma}U^\alpha \xi^\beta U^\gamma
\end{equation}
\noindent $\xi^\mu$ is the deviation vector and $U^\mu$ denotes four-velocity of a set of timelike observers following a geodesic. We consider $U^\mu=(1,0,0,0)$ and $\xi^u=0$. The deviation equations along $x$ and $y$ directions become ($A_+\neq0, A_\times=0$)
\begin{gather}
    \ddot{\xi}^{x}=-\frac{1}{2}A_+(u)\xi^x \label{eq:xi_x}\\
    \ddot{\xi}^{y}=\frac{1}{2}A_+(u)\xi^y. \label{eq:xi_y}
\end{gather}

\noindent Eqs.(\ref{eq:xi_x}) and (\ref{eq:xi_y}) are similar to the geodesic equations (\ref{eq:x_sqp}) and (\ref{eq:y_sqp}). Physically, $\xi^x$ and $\xi^y$ correspond to the separation between a pair of geodesics along $x(u)$ and $y(u)$ directions. Hence, considering two geodesic solutions \{$x_1,y_1$\} and \{$x_2,y_2$\}, the deviation vector is,
$$\xi^x=x_1-x_2, \hspace{1cm} \xi^y=y_1-y_2.$$ 
\noindent Considering $x_2=y_2=0$, we find that the solution $\xi^x=x_1(=x), \xi^y= y_2 (=y)$.
Thus, the deviation analysis would reveal identical results as obtained from solving the geodesic equations. However, both the methods differ for spacetimes having nonzero background curvature. In our past work \citep{Siddhant:2020,Chakraborty:2021}, we have shown that memory effects obtained using geodesics is equivalent to the total deviation (wave and background). Since in case of exact plane gravitational waves the curvature disturbance propagates over a flat spacetime, there is zero background deviation. Hence, the results are identical in both the methods. \\

\noindent In this article, our entire analysis relies on the choice of a pulse profile in the Brinkmann gauge. Hence, one may arrive at the conclusion that the entire analysis is {\em coordinate dependent} and gravitational memory, as defined in the present context, is not a gauge invariant observable and different from existing works in astrophysical settings. But, considering detectors (as in LIGO or LISA) that follow timelike geodesic paths, one can show that the final change in separation (memory) in Fermi normal coordinates takes an exactly similar form as given in Eqs.(\ref{eq:xi_x}) and (\ref{eq:xi_y}). This is because gravitational memory is imprinted in the {\em timelike Penrose limit} of the gravitational wave spacetime \citep{Shore:2018}. Since in the Penrose limit, one obtains the exact plane wave metric, the following equations obeyed by the test-detectors is similar to the ones given in Eqs.(\ref{eq:xi_x}) and (\ref{eq:xi_y}). In case of other spacetimes containing gravitational waves like Kundt wave geometries, one needs to analyse the equations in Fermi normal coordinates as seen by an idealized detector. Such treatment has been done in our earlier work \citep{Siddhant:2020, Chakraborty:2021}.

\section{Conclusions}

\noindent Our main emphasis here has been to present a simple 
analytical example of gravitational wave memory using a square pulse. To this end, we show how the nature of the pulse profile ({\em i.e.} the chosen functional forms of $A_+(u)$ and/or $A_\times(u)$) influences geodesic evolution in the exact plane gravitational wave spacetime, thereby leading to memory effects. 

\noindent We have discussed the two standard coordinate systems used while studying exact plane waves. The Brinkmann coordinates are employed in our calculations since they are free of {\em coordinate singularities}. These coordinates have an unconstrained metric function $H(u,x,y)$ denoting the profile and the polarization of the gravitational wave. We choose a {\em square pulse profile} (Fig.(\ref{fig:Squarepulse})) which represents a sandwich wave spacetime geometry.

\noindent In this square pulse geometry, we have a region containing plane gravitational waves (Region-II) sandwiched between two flat Minkowski regions (Region-I and Region-III). Setting initial transverse coordinate velocity of a pair of geodesics to zero in Region-I, we solve the geodesic equations analytically along $x$ and $y$ directions assuming that the solutions are continuous and differentiable (at least $C^1$) at the boundaries of the pulse ($u=-a,+a$).  The geodesic solution for the coordinate $v(u)$ is obtained once $x(u)$ and $y(u)$ is known (Eq.(\ref{eq:v_soln})).     

\noindent The analytical solutions obtained  are then illustrated through plots showing the  geodesic evolution in all the three regions of the spacetime. We analyse three distinct cases of linear polarization in this article viz; plus, cross and both (plus and cross combined). Our results show {\em monotonically increasing displacement memory} and {\em constant shift velocity memory} along $x$ and $y$ directions in all these cases. In the $v$-direction, we observe that the solutions are continuous but suffer from derivative discontinuity at the boundaries of the pulse owing to the {\em step-function} nature of the pulse profile (Eqs.(\ref{eq:pluspulse}) and (\ref{eq:first_integral_v})).

\noindent {In the first case where $A_+\neq0, A_\times=0$, we find geodesics meet along the $x$ direction. The focusing $u$-value is independent of the initial position but depends on $a$ and $A_0$. For the second case ($A_+=0, A_\times\neq0$), we get focusing which depends both on the geometry of the pulse and the initial positions along $X$ and $Y$ directions (in {\em normal coordinates}).  Finally, when $A_+=A_\times\neq0$, we find similar results as found in the case of cross polarization. Such meeting of trajectories signify the presence of benign caustics. A study of memory effects using geodesic congruences (also known as ${\cal B}$-memory) by the current authors \citep{Chak:2020} in this spacetime, have also shown the existence of such caustics.

\noindent Further, we have tried to demonstrate how the gravitational wave pulse distorts an initial configuration of a ring of particles. The loci of the configuration changes from a circle to an ellipse after the passage of the pulse. In case of plus polarization, we find the ellipse degenerates to a straight line along the $y$-axis. 
The focusing is dependent on the shape of the pulse ($\xi$). The expansion variable corresponding to this deformation at focusing diverges to negative infinity.  This is in agreement with the result from geodesic analysis where trajectories meet along the $x$-direction. 
 In the other case having only cross polarization, we find that the final configuration is an ellipse rotated by an angle of $\pi/4$ w. r. t. $x$-axis. Finally, we observe that the nature of shear is determined by the type of polarization of the gravitational wave ({\em i.e.} $\sigma_+\neq0,\sigma_\times=0$ for $A_+\neq0,A\times=0$ and vice versa). 

 }

\noindent Gravitational memory obtained from geodesic deviation is shown to yield identical results as obtained from the geodesic analysis. This happens because the background over which the plane wave propagates is flat and the entire contribution of the deviation comes from the radiation itself.

\noindent In conclusion, this article shows the crucial role played by 
explicit solutions of the geodesic equations in
developing analytical examples and understanding of 
gravitational memory effects, in exact plane gravitational wave spacetimes. Similar analytical exercises for other radiative geometries, if possible,
will surely be useful and may also 
reveal new features of memory effects as well as 
characteristics of the background spacetime.

\section*{Acknowledgments}

 \noindent I. C. is supported by University Grants Commission, Government of India through a Senior Research Fellowship with Reference ID: 523711.\\

\noindent {\bf Data Availability Statement:} This manuscript has no associated data
or the data will not be deposited.
\bibliographystyle{apsrev4-2}
\bibliography{mybibliography_sqp}

\end{document}